\documentclass[%
superscriptaddress,
twocolumn,
 amsmath,amssymb,
prl,
]{revtex4-2}

\usepackage{graphicx}% Include figure files
\usepackage{dcolumn}% Align table columns on decimal point
\usepackage{bm}% bold math
\usepackage{xcolor}
\usepackage[colorlinks]{hyperref}% add hypertext capabilities
\usepackage{orcidlink}
\usepackage{lineno}
\usepackage{aas_macros}

\begin{document}

% \preprint{APS/123-QED}

\title{The role of magnetar transient activity in time-domain and multimessenger astronomy}

\author{Michela Negro\,\orcidlink{0000-0002-6548-5622}}
\email{michelanegro@lsu.edu}
\affiliation{Department of Physics \& Astronomy, Louisiana State University, Baton Rouge, LA 70803, USA}

\author{George Younes\,\orcidlink{0000-0002-7991-028X}}
\affiliation{NASA Goddard Space Flight Center, 8800 Greenbelt Road, Greenbelt, MD 20771, USA}%

\author{Zorawar Wadiasingh\,\orcidlink{0000-0002-9249-0515}}
\affiliation{NASA Goddard Space Flight Center, 8800 Greenbelt Road, Greenbelt, MD 20771, USA}
\affiliation{Department of Astronomy, University of Maryland, College Park, MD 20742, USA}
\affiliation{Center for Research and Exploration in Space Science and Technology, NASA/GSFC, Greenbelt, MD 20771, USA}

\author{Eric Burns\,\orcidlink{0000-0002-2942-3379}}
\affiliation{Department of Physics \& Astronomy, Louisiana State University, Baton Rouge, LA 70803, USA}

\author{Aaron Trigg\,\orcidlink{0009-0006-8598-728X}}
\affiliation{Department of Physics \& Astronomy, Louisiana State University, Baton Rouge, LA 70803, USA}

\author{Matthew Baring\,\orcidlink{0000-0003-4433-1365}}
\affiliation{Department of Physics and Astronomy MS 108, Rice University, 6100 Main Street, Houston, TX 77251-1892, USA}

\date{\today}% It is always \today, today,

\begin{abstract}
Time-domain and multimessenger astronomy (TDAMM) involves the study of transient and time-variable phenomena across various wavelengths and messengers. The Astro2020 Decadal Survey has identified TDAMM as the top priority for NASA in this decade, emphasizing its crucial role in advancing our understanding of the universe and driving new discoveries in astrophysics.\\
The TDAMM community has come together to provide further guidance to funding agencies, aiming to define a clear path toward optimizing scientific returns in this research domain. This encompasses not only astronomy but also fundamental physics, offering insights into gravity properties, the formation of heavy elements, the equation of state of dense matter, and quantum effects associated with extreme magnetic fields. Magnetars, neutron stars with the strongest magnetic fields known in the universe, play a critical role in this context.\\
In this manuscript, we aim to underscore the significance of magnetars in TDAMM, highlighting the necessity of ensuring observational continuity, addressing current limitations, and outlining essential requirements to expand our knowledge in this field.
\end{abstract}

\maketitle

%\tableofcontents

% \textcolor{red}{PRL	Letter: 3750 words (current draft ~3653), Comment/Reply: 750 words}

Over the last two decades, magnetars have been the subject of numerous comprehensive review articles. Notable among these, Mereghetti's 2008 work delved into observational evidence distinguishing a unique class of isolated neutron stars ---powered by magnetic energy--- termed magnetars, which encompass anomalous X-ray pulsars (AXP) and soft gamma-ray repeaters (SGR) \citep{2015SSRv..191..315M}. Subsequently, in 2015, Mereghetti, alongside Pons and Melatos, offered a second review focusing on persistent emission properties, exploring models explaining extreme magnetic field origins, evolutionary pathways, and interconnections with other neutron star classifications \citep{2015SSRv..191..315M}. Additionally, in the same year, Turolla, Zane, and Watts \citep{2015RPPh...78k6901T} provided a detailed overview of magnetar origins and evolution, emphasizing theoretical modeling's critical role in understanding fundamental physics, constrained by both persistent and transient emission observations. Furthermore, Kaspi and Beloborodov's 2017 review \citep{kaspi2017}, and four years later Esposito, Rea, and Israel's 2021 contribution \citep{esposito2021ASSL}, updated the discourse on the magnetar population within our Galaxy. These reviews focused on high-energy (X-rays and above) persistent emission characteristics, temporal behavior, and transient activities, collectively enriching our understanding of these enigmatic celestial objects. Recently, Dell'Orso and Stella provided a review focused on newly-born millisecond magnetars \citep{2022ASSL..465..245D}. 

\begin{figure*}
    \centering
    \includegraphics[width=\textwidth]{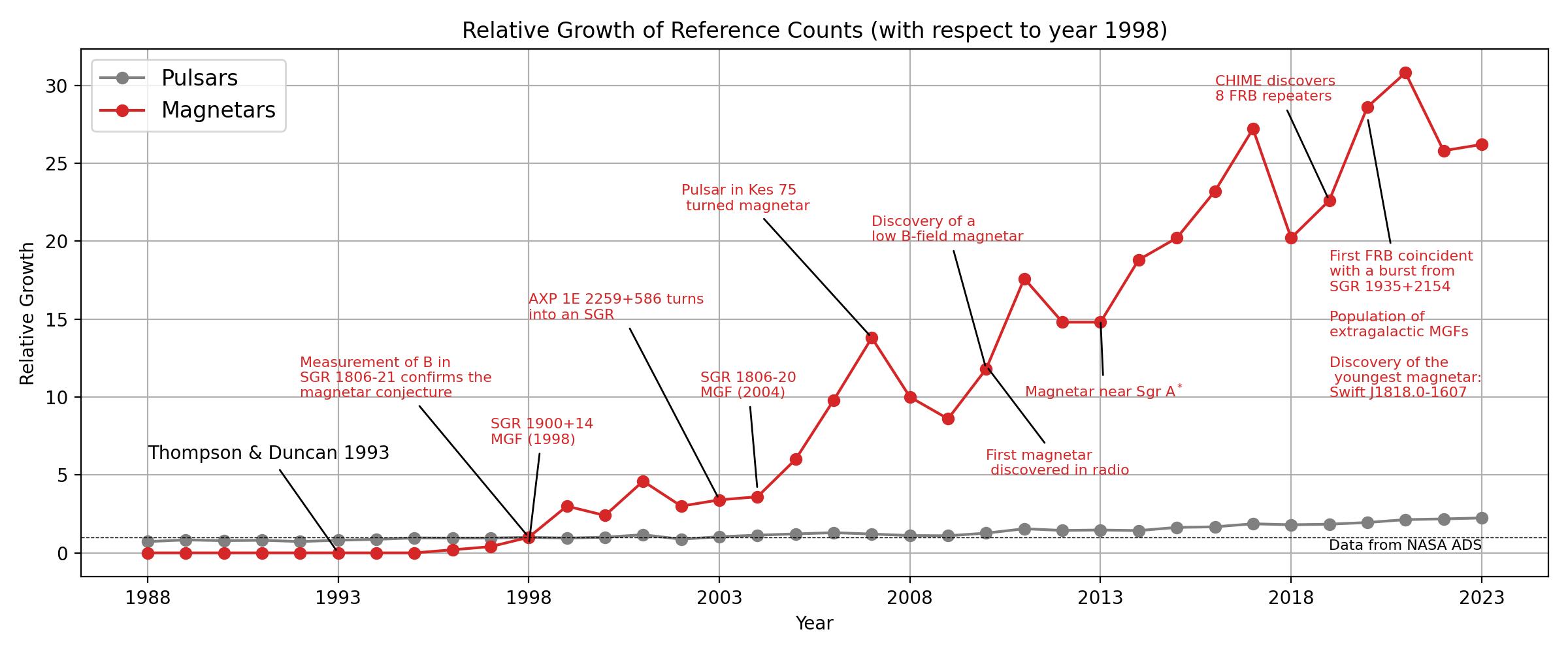}
    \caption{Relative growth of peer-reviewed articles mentioning ``magnetars'' compared to articles growth mentioning ``pulsars'' normalized to 1998. In red annotations we highlight some major events and observations related to magnetars.
    % \textcolor{red}{list here all the refs for the flags in the plot}; Detection of 8 FRB repeaters \cite{2019ApJ...885L..24C}; discovery of low B-field magnetar \cite{2010Sci...330..944R}
    }
    \label{fig:papers}
\end{figure*}

A clear trend emerging from each of these reviews is that, despite representing only a small fraction of the observed neutron star population, magnetars have been attracting the interest of an increasing number of scientists from many different area of astronomy and astrophysics, demonstrated by the relative growth of publications mentioning ``magnetars'' over the past decades (Figure~\ref{fig:papers}). The sheer number of reviews is the result of continued fundamental discoveries within the magnetar field that shape the understanding of the NS population at large and beyond. Chief among them is the unification of AXP and SGR under the same name \citep{duncan1998global}, and progresses with the phenomenology of starquakes\citep{1996Natur.382..518C}, the observation of the extremely bright events called giant flares (when a magnetar outshines the Sun for a fraction of a second in hard X-rays) \citep[e.g.,][among others]{hurley1999giant, hurley2005exceptionally}, the identification of a population of extragalactic magnetars' flares masquerading as short gamma-ray bursts, the observation of new mysterious bright and intermittent Galactic long-period radio pulsating sources, and last but not least the connection between magnetars and fast-radio bursts \citep{mereghetti20ApJ}. Such type of observational evidence has been catalyzing studies and increasing interest of a wider and deeper community. Several models predict gravitational waves emission from magnetars at birth and during giant flares, and many theoretical studies suggest that high-energy neutrinos should be expected during those events. Furthermore, there are discussions about magnetars being behind other types of isolated neutron stars, such as, to name a couple, central compact objects (CCOs) and X-ray dim isolated neutron stars (XDINs), but also other types of transient events ultra-luminous X-ray sources (ULXs), super-luminous supernovae (SLNS), and fast X-ray transients (FXT).

This manuscript seeks to explore the significant role of magnetars in time-domain and multimessenger (TDAMM) astronomy, focusing therefore on their transient activity. Core aspect of this scope is highlighting the critical role of the high-energy space-based missions that have enabled inference built upon compelling evidence in the past several decades.  Our aim is to highlight the main characteristics of these missions while also acknowledging their limitations, thereby proposing viable avenues for enhancing our capacity to study these captivating celestial entities. In Section~\ref{sec:1} we give a brief introduction on the magnetar population, while focusing on the fast-transient activity in Section~\ref{sec:2}, where we discuss on short bursts, storms, and flares. In Section~\ref{sec:3} we delve into the multimessenger prospects for magnetars discussing expectations for the observation of gravitational waves and neutrinos.

\section{Magnetars} 
\label{sec:1}
The question of what conditions are necessary to create and power a magnetar underpins the broad interest in these enigmatic celestial objects. To grasp the significance of this question, it's essential to first define what a magnetar is. Neutron stars (NS) are the compact remnants forged in the explosion of massive stars during a supernova event.  With a mass typically ranging between about 1.2 and 2 times that of the Sun, the neutron-degenerate matter in NS is squeezed into a sphere only about 10 to 20 kilometers (6 to 12 miles) in diameter, reaching supra-nuclear densities in their interior and representing the densest form of matter known in the universe ---about fourteen orders of magnitude denser than osmium, the densest element found on Earth. NS are highly magnetized, which requires a dynamo-like amplification of an original magnetic field from stellar mergers \citep{2019Natur.574..211S}, fall-back dynamos \citep{2022A&A...668A..79B} or other mechanisms, like inverse cascading of helical and fractionally helical magnetic fields\citep{2020ApJ...901...18B}. Differential rotation and rotation-convection coupling in the collapsing core of massive stars can also start this dynamo effect \citep{duncan1992formation}, which continues in the convective inner structure of the newly formed rapidly rotating NS, and produce strong magnetic fields \citep{TD1993}. Magnetars are ultra-magnetized NS, with recorded (dipolar) magnetic fields of the order of $10^{14-15}$ G, usually found in isolation (i.e. not in binary systems) and sometimes associated to a nearby supernova remnant. Understanding what type of progenitor star(s) can generate magnetars is key to understand their nature and behaviour, which ultimately give access to the physical mechanisms involved in such extreme environments. 

\begin{figure*}
    \centering
    \includegraphics[width=\textwidth]{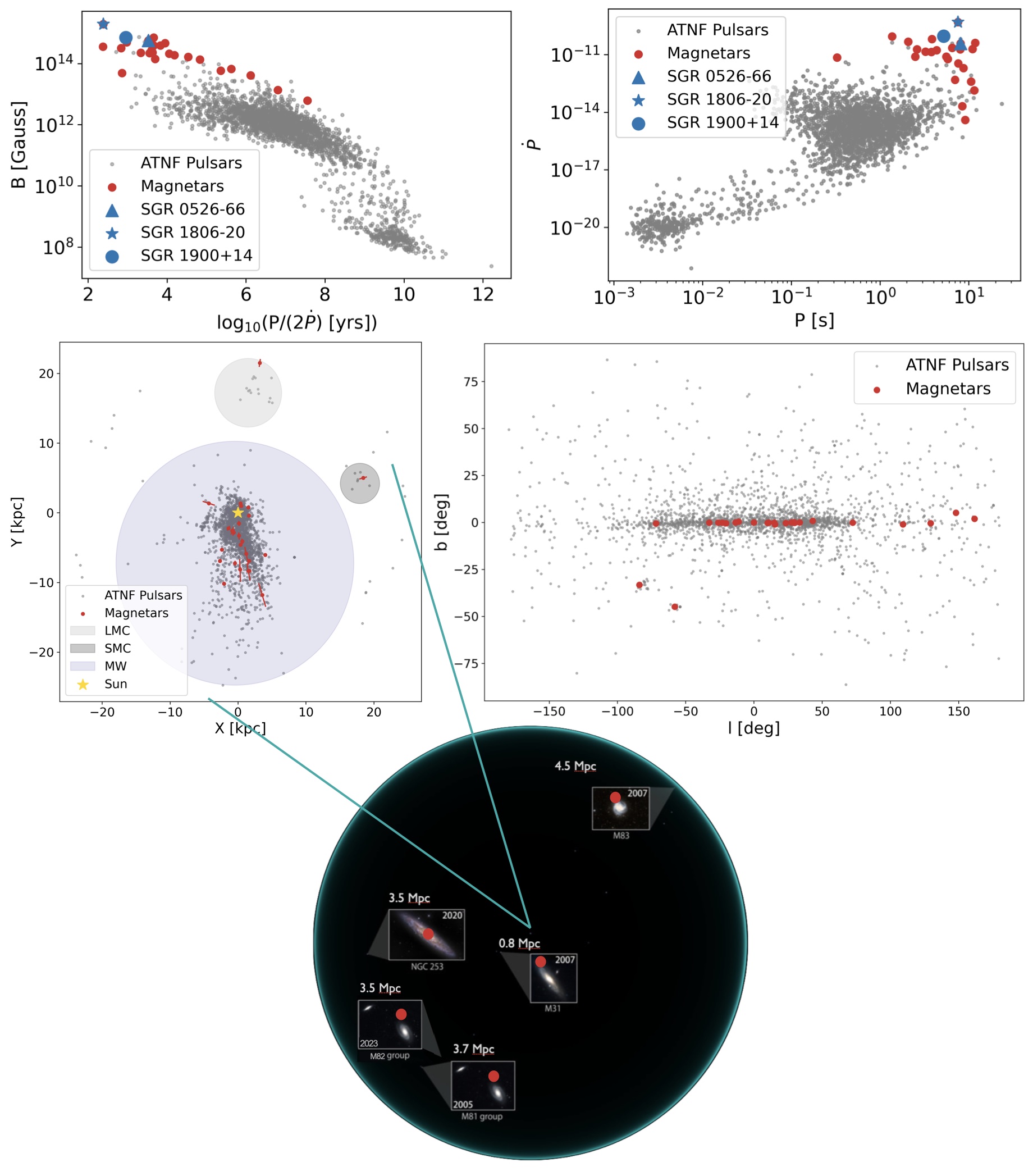}
    \caption{The population of Galactic high-energy magnetars in red (data from the McGill catalog \citep{2014ApJS..212....6O}, compared to the pulsar population provided by the Australia Telescope National Facility (ATNF) catalog \citep{2005AJ....129.1993M} shown in grey. Top left: Shows the magnetic field strength as a function of the estimated age; Top right: the spin-down rate as a function of the period. Middle: Top-Galactic view (left) and mollweide sky-projection (right) showing the distribution in the Milky way. Bottom:  population of known extragalactic magnetar candidates (Credit: adapted from NASA Goddard press release).}
    \label{fig:Gal_population}
\end{figure*}

Studying the population of magnetars in our Galaxy, both alone and in comparison to the bigger population of isolated NSs, can shed light on their progenitors \citep{Beniamini2019}, and hence the conditions necessary for their formation. The main characteristics of the  Galactic magnetar population is illustrated in Figure~\ref{fig:Gal_population}, in comparison to the wider pulsars population \citep[ATNF Catalog]{2005AJ....129.1993M}. About thirty known high-energy emitting magnetars are hosted in our Galaxy, for the majority located within 1 degrees from the Galactic plane, one in the Large Magellanic Cloud \citep{cline1981precise} and one in the Small Magellanic Cloud \citep{2002ApJ...574L..29L}. They are characterized by a high spin-down rate and slow rotation period, which together with their location in the plane, suggests that active magnetars are typically young, from approximately a hundred years (the youngest known is about 240 years old, discovered in 2020 \citep{2020ApJ...896L..30E}) to a few tens of thousands of years, as shown in the top-left panel of Figure~\ref{fig:Gal_population} in comparison with the much older population of pulsars.  Except for PSR~J1622-4950, which was discovered in radio in 2010 \citep{levin10ApJ}, all the other known ``standard" young magnetars revealed themselves in the X-ray band, most of them through a first bright transient event. Through the observation and statistical study of the bursting activity of SGR~1806-20, Cheng et al in 1995 \cite{1996Natur.382..518C} found evidence of the hypothesized solid crust on magnetars surface \citep[see for a similar association with FRBs,][]{2023MNRAS.526.2795T}. In fact, they found similarities between the magnetar bursts energy and waiting time distributions and those for quakes on Earth caused by tectonic movements \citep{perna11ApJ, 2020ApJ...902L..32D}.  The discovery of fast radio bursts (FRBs) from Galactic magnetars \citep{bochenek2020Nat, chime20Natur} in coincidence with their bursting   \citep{mereghetti20ApJ, liNatAs1935} and possibly glitching activity \citep[e.g.,][]{younes23NatAs, ge24RAA1935} have been providing crucial information on the physical mechanisms that power these phenomena, and the crustal and magnetospheric conditions that can produce FRBs. More recently, radio transient surveys have been discovering a population of long-period Galactic radio pulsars which are likely older magnetars \citep{2022NatAs...6..828C,2022Natur.601..526H,2023Natur.619..487H,2022ApJ...940...72R,2023MNRAS.520.1872B,2024ApJ...961..214R}. None of these have a high-energy counterpart yet, but given their highly-variable nature in the radio and likely magnetar nature along with possible connection to long-period FRBs \citep{2020MNRAS.496.3390B}, it is plausible future X-ray and gamma-ray transients could be associated with high-energy monitors with sufficient angular resolution.

Several magnetar candidates have been pinpointed in neighboring galaxies: NGC 253 (Sculptor galaxy), M31 (Andromeda galaxy), the M81-M82 group, and M83. Confirming pulsating emissions matching typical magnetar rotation periods would validate their identity. Extragalactic magnetars can be observed only during the brightest flares, but unequivocal association requires the detection of the pulsating emission, typically too fast-fading to be caught in time by sensitive instruments. However, the detection of the brighter short initial spike allows to infer the volumetric intrinsic rates of such phenomena associated to magnetars, providing important clues on their formation channels \citep{burns2021identification}. The current population of extragalactic magnetar candidates counts only a handful of objects from nearby galaxies, limiting our constraints on the intrinsic rates. Such limitation needs more sensitive all-sky soft gamma-ray monitors able to trigger more efficiently on these events.  
% \textcolor{red}{Other important inference that have been made through the observation of magnetars transient activity? possible relevant for the physics involved?}

\section{Magnetars in Time Domain Astronomy}
\label{sec:2}
\begin{figure*}[t]
    \centering
    \includegraphics[width=\textwidth]{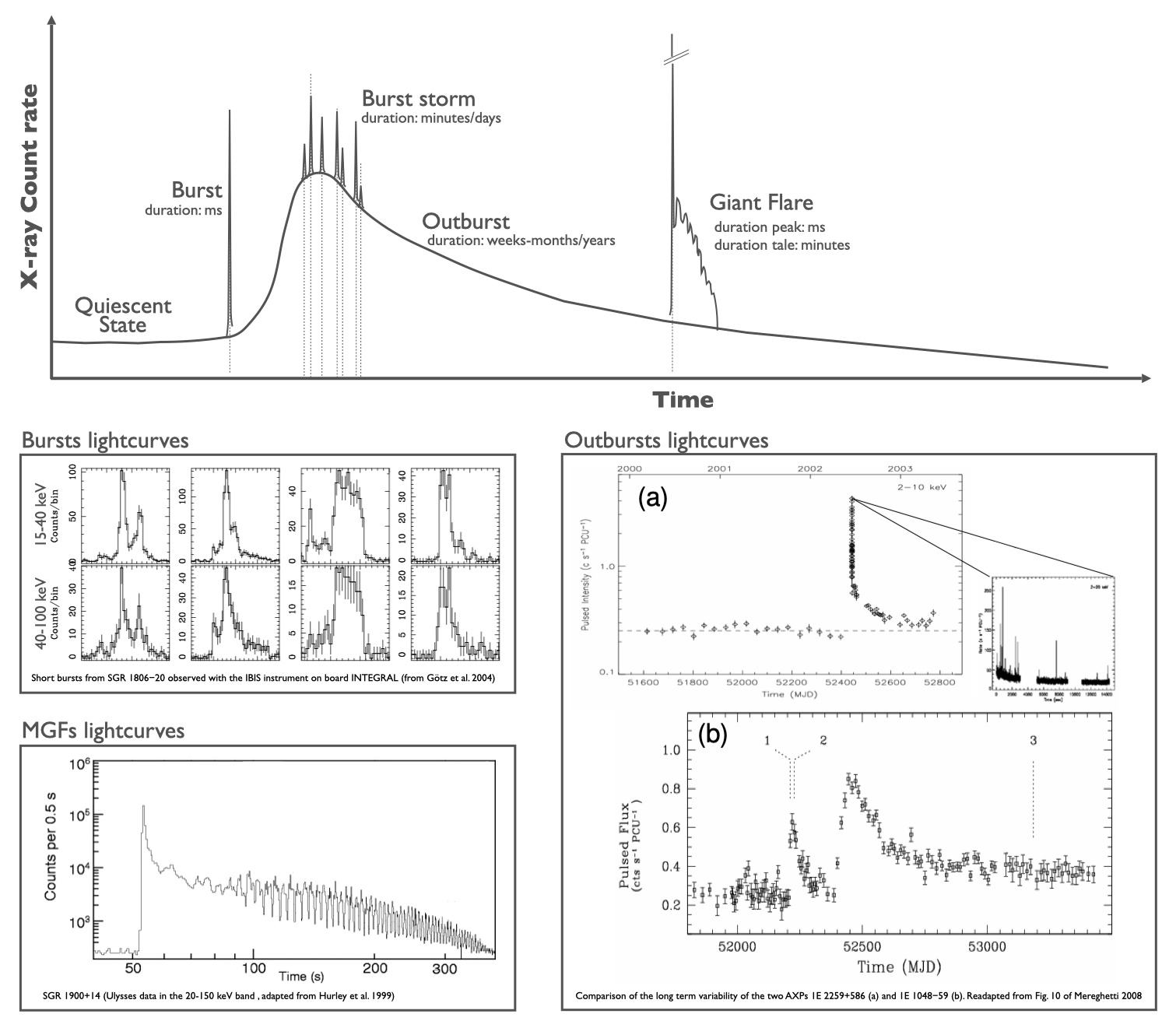}
    \caption{Illustration of the topology of transient events from magnetars. Top panel summarizes the different transient activities of magnetars. The panel 'Bursts lightcurves' shows examples of short bursts from SGR~1806-20 observed by INTEGRAL-IBIS in 2004 \citep{2004A&A...417L..45G}. The panel 'MGFs lightcurves' shows the 1998 MGF from SGR~1900+14 as representative of this class of events \citep{hurley1999giant}. The panel 'Outbursts lightcurves' shows a comparison of the long-term variability of the two AXPs 1E~1048-59 (b) \citep[adapted from Fig. 10 of][]{2015SSRv..191..315M}. }
    \label{fig:magnetrasActivity}
\end{figure*}
Magnetars show a variety of transient activity observable from soft X-rays up to medium energy gamma-ray bands and differring in terms of timescales, energetics and temporal and spectral evolution. Such activity includes outbursts, short bursts, burst storms, and flares as illustrated in (Figure~\ref{fig:magnetrasActivity}). 

Apart from the more prolonged outbursts, magnetars are also associated with short bursts of intense radiation, typically lasting only a fraction of a second. These short bursts, often observed in the hard X-ray spectrum, provide valuable insights into the extreme physical conditions prevailing in the vicinity of magnetars. The origins of these short bursts may be linked to the sudden release of magnetic energy, or magneto-elastic energy from the crust. Short bursts are quasi-thermal, and broadband soft gamma-ray spectroscopy reveals they are consistent with trapped fireballs within closed loops at low altitudes in the magnetosphere \citep[e.g.,][]{2012ApJ...749..122V,2014ApJ...785...52Y}.

Magnetars are known for their outbursts, during which the quiescent/persistent X-rays emission increases by as many as 3 orders of magnitude. Typically these events are characterized by a faster (hours - days) flux rise, followed by a slower (weeks- to years-long) decay to return eventually to quiescence. Such temporal characteristics enable follow-ups and monitoring by sensitive pointing telescopes, providing accurate flux estimates. Crustal shifts due to magnetic stress are believed to be the cause of magnetars outbursts. Short bursts and flares have been observed during outbursts, as well as isolated in time. Burst storms have been observed to happen at the onset of outbursts. %Depending on the total energy released, the bursts can be divided into three categories: bursts (isotropic energies, $E_{iso} \leq 10^{41}$ ergs), and flares ($E_{iso} > 10^{41}$ ergs))

Magnetar burst storms (or burst forests) refer to periods of heightened and sustained activity, during which a magnetar emits a series of tens to thousands of bursts over a relatively short time frame of minutes to days. These stormy episodes contribute significantly to our understanding of the magnetar's dynamic behavior. Studying these storms helps decipher the underlying processes that govern the interplay between the decaying intense magnetic field, the internal and external structure of the magnetar, and the radiative processes occurring in high-B-field regime close to the surface of the magnetar.

Magnetar flares represent another facet of their transient activity, characterized by sudden and intense increases in radiation across multiple wavelengths. Such events are characterized by a ms-long bright spike, followed by a dimmer (but still bright) periodic tail decaying in time. These flares are among the most energetic events in the universe, releasing energy on the order of solar flares but with magnitudes far surpassing them \cite{hurley1999giant, hurley2005exceptionally, 2008ApJ...685.1114I}. Typically classified in intermediates flares (with $E_{iso} \sim 10^{41}-10^{43}$ ergs, and giant flares (MGFs) ($E_{iso} \sim 10^{44}-10^{47}$ ergs), magnetar flares are crucial to enhance our comprehension of the extreme conditions prevailing in the vicinity of these celestial bodies and provides valuable data for refining models of magnetar behavior.

\begin{figure*}
    \centering
    \includegraphics[width=\textwidth]{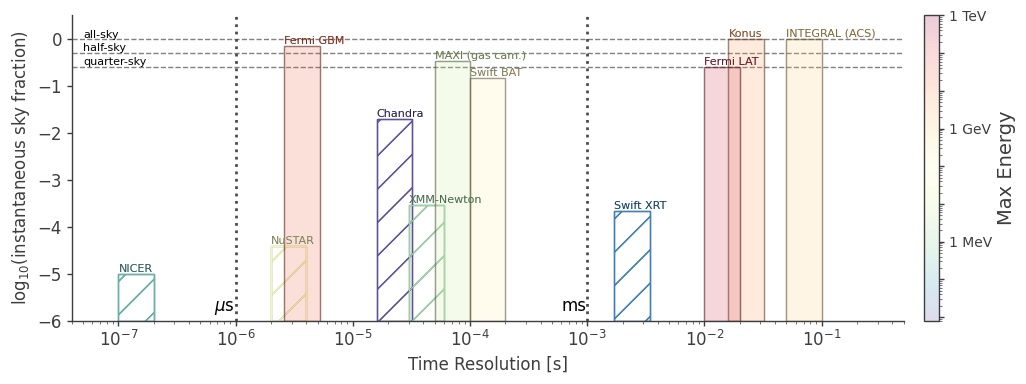}
    \caption{Active high-energy missions available for magnetar transient activity observations. On the y-axis we show the sky coverage, monitors are marked with filled rectangles, while pointing telescopes with hatched rectangles. On the x-axis we report the time resolution of the different instruments. The colors mark the upper end of the energy range covered by the instruments.} 
    % \textcolor{red}{to be changed into energy vs time resolution and add reference for busts outbursts and flares}}
    \label{fig:missions}
\end{figure*}

High-energy monitors, spectrometers and fast-repointing instruments have been enabling the observation of magnetars' dynamic transient activity since the '80s. Past major contributors include ROSAT \citep{1982AdSpR...2d.241T}, CGRO \citep{1994ApJS...92..351G}, RXTE \citep{1999NuPhS..69...12S}, and BeppoSAX \citep{1997A&AS..122..299B}. In Table~\ref{tab:instruments} we list the major high-energy instruments that are currently contributing to monitoring and detection of magnetars' transients activity. High-energy instruments like the GBM on board Fermi, Konus on board of WIND, BAT on board of Swift and the ACS on INTEGRAL, offer broad coverage of the soft gamma-ray band, making them valuable for detecting a wide range of transient events, including those from magnetars. Instruments like Chandra, XMM-Newton, NuSTAR, and NICER provide high-angular resolution and are capable of pinpointing the precise locations of transient events, aiding in follow-up studies and multi-wavelength observations, however, except for Swift with minute-scale reaction, the repointing time limits follow-ups to magnetars outbursts and burst storms. 
Figure ~\ref{fig:missions} is a visual illustration of the available energy, timing and sky coverage provided by the instruments listed in Table~\ref{tab:instruments}.

\begin{table*}[t]
  \centering
   \begin{tabular}{llclll}
  \toprule
 {\small \textbf{Mission}}   & {\small\textbf{FoV}} & {\small\textbf{Min. Repoint}} & {\small\textbf{Energy (keV)}} & {\small\textbf{Time Res.$^+$}}  & {\small\textbf{Launch}}\\ 
 \hline
    % CGRO & Pointed & 1 & 10-10000 & ... & 1991-1999 & Ref 1 \\
    % BeppoSAX & Pointed & 1 & 0.1-300 & ... & 1996-2003 & Ref 2 \\
    % ROSAT & Pointed & 3600 & 0.1-2.4 & ... & 1990-1999 & Ref 3 \\
    % RXTE & Pointed & 0.1 & 2-200 & ... & 1995-2012 & Ref 4 \\
    Konus-WIND & all-sky & -- & $20-20000$ & $16$ ms & 1994  \\
    Chandra (HRC) & $30'\times30'$ & $<5$ days & $0.1-10$ & $16\mu$s & 1999 \\
    XMM-Newton (PN) & $27.5'\times27.5'$ & $<24$ hours & $0.2-12$ & 30 $\mu$s & 1999\\
    INTEGRAL (ACS) & all-sky & -- & $>80$ & $50$ ms & 2002\\
    Swift BAT & $15$\% & -- & $15-350$ & $100$ $\mu$s & 2004 \\
    Swift XRT & $23.6'\times23.6'$ & minutes & $0.3-10$ & $1.7$ ms & 2004\\
    Fermi GBM & $70$\% of sky & -- & $8-40000$ & $2.6$ $\mu$s & 2008 \\
    Fermi LAT & $25$\% of sky & -- & $(0.1-800)\times10^{6}$ & $10$ ms & 2008\\
    MAXI (gas cam.) & $1.5^\circ\times160^\circ$ & -- & $2-30$ & $50$ $\mu$s & 2008 \\
    NuSTAR & $10'^*$ & $<24$ hours & $3-79$ & 2 $\mu$s & 2012 \\
    NICER & $5'^{**}$ & $<4$ hours & $0.2-12$ & $100$ ns & 2017 \\
  \end{tabular}
   \caption{Major high-energy instruments currently contributing to magnetar observations. Missions figures of merit can be found on the NASA's High-Energy Astrophysics Science Archive Research Center (HEASARC). For the time property of INTEGRAL anti-coincidence shield (ACS) we referred to Ref.~\citep{2012A&A...541A.122S}.\\ $^{*}$ FoV (50\% resp.) at 10 keV. \\$^{**}$Non-imaging.\\ $^{+}$ This is the highest temporal resolution reached in any mode of any instrument on-board.}
   \label{tab:instruments}
\end{table*}

\subsection{Outbursts}
Most magnetars display periods of elevated X-ray emission above their historical minimum level, sometimes by as many as 3 orders of magnitude in luminosity\citep{zelati18MNRAS}, i.e., magnetar outbursts. These epochs, which are typically observed concurrently to the onset of bursting activity (see below), are defined by extreme spectral-temporal variability to the soft and hard X-ray emission in the form of harder spectra, pulse profile and fraction variation, timing noise and glitches \citep[e.g.,][]{Gavriil+04, Woods2007, rea09MNRAS:0501, dib14ApJ, Hu2020, younes22ApJ:1830}. At radio wavelengths, six confirmed magnetars have shown transient {\sl radio pulsed emission}, appearing around outburst epochs (\cite{camilo06Natur, lower20ApJ:1818}, or, in a few occasions, disappearing, e.g., \citep{lower23ApJ}). For a few magnetars, the infra-red to optical emission have also been observed to vary \citep{tam04ApJ:2259}. These outburst epochs last anywhere from months to years during which the multi-wavelength properties usually return back to their pre-outburst states (see, e.g. \citep{zelati18MNRAS}, for a review).

Magnetar outbursts are generally attributed to crustal shifts (e.g. due to stresses on the surface from internal B-field restructuring and perhaps decay), imparting a twist onto an external magnetic field loop (see \citep{2015RPPh...78k6901T} and references therein). The surface heating arises due to either energy deposition in the crust, e.g., from Hall wave avalanches \citep{thompson96ApJ, beloborodov16ApJ}, or bombardment of the surface by accelerated particles in a twisted external B-field \citep{beloborodov09ApJ}. Both models predict the formation of surface hot-spots, which could explain the altered pulse shape and amplitude during magnetar outbursts as well as the harder spectra and increased X-ray power. Although in both cases the outburst is initiated by an elastic failure of the crust \citep{2020ApJ...902L..32D}, their evolution is dictated by different regions of the NS. For the external model, as the twisted fields ``unwind'', magnetic energy is released in the form of radiation, typically leading to the shrinkage and cooling of the hot-spot \citep{beloborodov09ApJ}, whereas if the heating is purely internal, the outburst decay is determined by crustal cooling scenarios heavily dependent on the micro- (e.g., crust impurity) and macro-physics (depth and total energy deposited in the crust\citep{brown09ApJ, pons09AA}).

Hence, given all the above, multiwavelength follow-up studies of magnetar outbursts are distinctly revealing of their highly dynamic nature; physics of plastic deformation of the crust, characteristics of the twisted B-field loops (twist magnitude, loop locale and total volume, etc.), pair-production and particle acceleration required for the emission of coherent radio emission, and the interconnection between all of these elements.

The high-energy properties of magnetar outbursts have been extensively studied with RXTE, XMM-Newton, Chandra, Swift/XRT, NuSTAR, and most recently NICER. Yet, the most consequential results have come from the long-term monitoring previously afforded by RXTE \citep{dib14ApJ} and currently conducted with XRT \citep[e.g.,][]{Archibald2013, Archibald2020} and NICER \citep{younes20ApJ:2259, lower2020}. Apart from the obvious benefit of such observational campaigns, i.e., the measurement of the period and period-derivative, hence, of the fundamental properties of the sources (magnetic dipole field strength, spin-down age, and spin-down power), continuous long-term monitoring of several bright magnetars from 1998 to 2012 revealed the common detection of some timing anomalies, mainly in the form of large spin-up (or on one occasion spin-down) glitches, at the onset of outbursts likely implying an internal trigger mechanism to these events \citep{dib14ApJ, Archibald2013}. Moreover, these monitoring campaigns revealed the delayed, erratic variability in the spin-down torque of these sources months to years after outburst onset providing clues to the dynamics of the untwisting magnetospheric B-field lines \citep{Woods2007, younes17ApJ:1806, Archibald2020}. Most recently, NICER (with the added benefit of the large effective area, relatively low background, and ease of repointing), through almost daily observations of the magnetar SGR 1830$-$0645, was able to resolve, for the first time, pulse peak migration which simplified the triple-peaked pulse profile at outburst onset to a single-peak in 37 days \citep{younes22ApJ:1830}. These results are the strongest evidence yet for plastic motion of the crust, long theorized to drive magnetar outbursts. Last but not least, for the same reasons, NICER has been able to time fainter magnetars, especially around periods of strong X-ray and radio bursting activity. Target of opportunity campaigns have been particularly revealing. A very recent example is provided by the FRB-emitting magnetar SGR 1935+2154, for which a double glitch event within 9 hours was detected, bracketing the largest spin-down rate ever observed from a NS along with an FRB \citep[][see also \citealp{younes23NatAs}]{Hu-2024Natur.626..500H}. This discovery has implications for the rate of superfluid material in a magnetar, outflowing plasma-loaded wind, production mechanism of FRBs in magnetars, and possibly gravitational wave emission.

Long-term monitoring of magnetars in X-rays (in tandem with radio and infrared campaigns) are unquestionably fruitful. In this regard, continued operation of Swift and, especially, NICER is essential, and looking into the future, the operation of a satellite with a similar type of capabilities such as Strobe-X.

\subsection{Short bursts and burst storms}
\label{sec:magnetars_bursts}
\begin{figure*}
    \centering
    \includegraphics[width=0.9\textwidth]{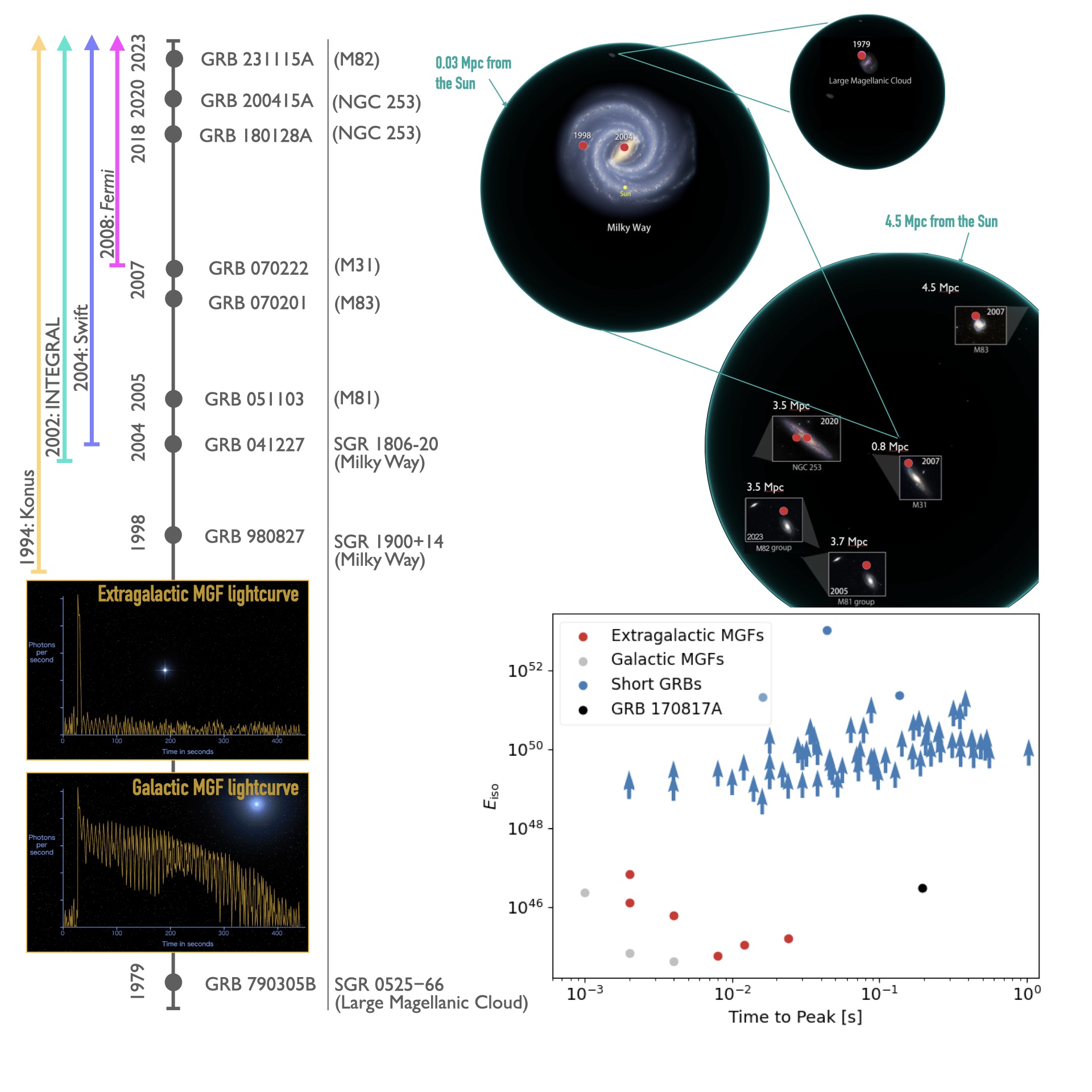}\\
    \includegraphics[width=0.9\textwidth]{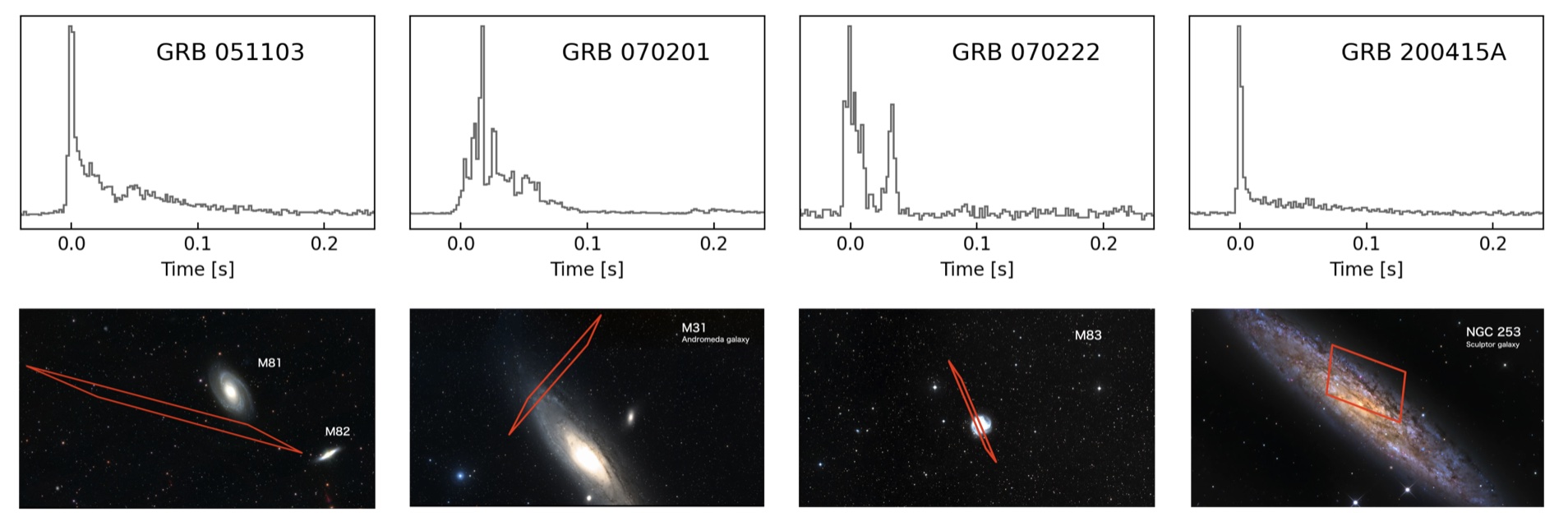}
    \caption{Illustration summarizing the current known population of MGFs and MGF candidates. The plot of the intrinsic energetic as a function of the rise time is the updated version of the plot presented in Ref.~\cite{2023IAUS..363..284N} with the addition of 2 more recent extragalactic identified events (see text for more details). We also report the lightcurves and the IPN localizations for four of the MGF candidates identified in Ref~\cite{burns2021identification}. (Credit images: adapted from NASA Goddard).}
    \label{fig:magnetrasTDAMM}
\end{figure*}

Short bursts are one of the most unique and defining property of the magnetar population. These sub-second, bright hard X-ray flashes, capable of reaching luminosities of about $10^{42}$~erg~s$^{-1}$ (Figure~\ref{fig:magnetrasTDAMM}), are easily identifiable by a suite of past and present large field-of-view hard X-ray monitors. They have played a crucial role in the inception of the soft gamma-ray repeater class \citep{atteia87ApJ, laros87ApJ, kouveliotou87ApJ}, and cementing the Anomolous X-ray pulsar (AXP) class as part of the same underlying population \citep{kaspi03ApJ}: NSs with activity driven by the extreme magnetic field strength \citep{Paczynski1992, duncan1992formation}. Magnetar short bursts can occur in isolation when one to few events are observed over the course of days, or, for the most active magnetars \citep[which tend to be the youngest,][]{perna11ApJ}, during burst storms/forests when hundreds to thousands are emitted over the course of minutes to hours \citep{collazzi15ApJS}.

Due to the dimness of most magnetars during their quiescent state, the large absorbing column in their direction (being at low Galactic latitudes), and the lack of adequate large field-of-view X-ray instruments\footnote{eROSITA might detect few magnetars at the end of its full-sky survey, yet these will likely be marked as candidates as many might not be bright enough for pulsation detection}, magnetar discoveries rarely occur through their persistent X-ray emission. This is plainly demonstrated through the discovery space of new magnetars in the last 20 years, which is fully dominated by the detection of short bursts, primarily with the Swift Burst Alert Telescope (BAT). BAT is sensitive to short magnetar bursts and able to localize them to within few arcminutes. The rapid follow-up with the Swift X-ray telescope (XRT) confirms the activity through the detection of the (at the time) bright X-ray counterpart, and provides arcsecond localization. Follow-up X-ray observations with the adequate time-resolution (which currently happens primarily with NICER), detects the pulse period of the source and its derivative, thus confirming the magnetar nature of the source \citep[see, e.g.,][among numerous ATels of this kind]{2020ATel14112....1R}. In summary, during its 20-year operation, the Swift telescope has enabled the discovery of more than double of the confirmed magnetar population in the Galaxy, and identified numerous new outbursts from the already known ones \citep[e.g.,][]{kaspi2017, esposito2021ASSL}. Chief among those discoveries are the identification of other classes of NSs as capable of showing magnetar-like activity, most noticeably high-B radio-pulsar \citep{archibald2016ApJ, gogus16ApJ}, CCOs \cite{rea2016ApJ}, and low-field magnetar\cite{rea2010Sci}, as well as the discovery of a canonical magnetar with a bright X-ray wind nebula \citep{younes2016ApJ}, a property typically reserved to rotation-powered pulsars \citep{kargaltsev2015SSRv}. These discoveries have enabled a more comprehensive understanding of what constitutes a magnetar, observationally, and theoretically, the latter through magneto-thermal evolutionary studies of poloidal and toroidal/crustal fields in NSs \citep{pons09AA, Vigano2013, gourgouliatos2016PNAS, 2020ApJ...903...40D, 2021NatAs...5..145I, 2023MNRAS.518.1222D, 2023MNRAS.523.5198D}.

Magnetar short bursts have also become crucial for the understanding of the enigmatic FRBs (see Section \ref{sec:magnetar_frb}). Following the detection of a short X-ray burst coincident with an FRB-like radio emission from the magnetar SGR 1935+2154 \citep{bochenek2020Nat, chime20Natur, mereghetti20ApJ, tavani21NatAs, liNatAs1935, ridnaia21NatAs}, comparison studies of the spectral and temporal properties of the FRB-associated X-ray short burst to those without an FRB counterpart (which is the overwhelming majority of short X-ray bursts) have shed light on the unusually hard spectrum of the X-ray burst that accompanied the FRB \citep{mereghetti20ApJ, younes2021NatAs}. This likely pointed to an active region in the vicinity of the open-field line zone which permits the release of bright radio waves away from the presumably dense environment of the closed magnetosphere \citep{younes2021NatAs}. Moreover, population-wide comparison of extragalactic FRBs and magnetar short bursts, such as duration, rate, waiting time distribution, etc. have shed some light onto the origin of extragalactic FRBs \citep{cruces21MNRAS, wei21ApJ}. Yet, these have not been able to confirm what fraction of the latter are indeed magnetars. This is partly due to our poor knowledge of the magnetar population in the Galaxy, and their activity cycle. Moreover, the detection of FRB 20200120E from a globular cluster in the nearby galaxy M~81 \cite{bhawdwaj21ApJ} challenges the notion that most, if not all, FRBs have a magnetar central engine, unless these magnetars were formed through unconventional channels, e.g., accretion induced collapse or the merger of two white-dwarfs. X-ray observations of this FRB 20200120E with current X-ray instruments ruled out coincident short bursts that are at the high end of the burst fluence distribution ($L_{\rm X}\gtrsim10^{42}$~erg~s$^{-1}$), and approaching the luminosities of intermediate flares \citep{pearlman23arXiv230810930P}.

Several advances in the magnetar field could be achieved with modest effort and investment. For instance, we currently lack a comprehensive, preferentially live, catalog of magnetar short bursts; an essential first step to understanding the activity rate and cycle of the population as a function of, e.g., spin-down age and magnetic field strengths. This could inform population studies of FRBs and comparison to the magnetar population. Ensuring the continued operation of Swift (or a new Swift-like instrument) is crucial for the continued discovery of new magnetars, and other exotic sources that exhibit magnetar-like activity. For instance, the low magnetic field magnetar, SGR 0418+5729, bares a striking resemblance during quiescence to XDINs \citep[][]{harbel07ApSS}. None of the latter sources (known as the magnificient seven) have shown magnetar-like activity, yet, this could be due to their larger ages, implying a lower rate of activity compared to canonical magnetars. If XDINs are confirmed to be magnetars, this would have significant consequences on the number density of magnetars in the Milky Way and their formation rate, providing clues for the birth process of magnetars \citep{Beniamini2019}. Looking into the future, new large FOV hard X-ray monitors that are capable of providing arcminute localization, preferentially equipped with sensitive follow-up X-ray instruments, e.g., NICER-like effective area, are key for the continued scientific success in our understanding of the magnetar bursting and outburst phenomena. Additionally, next generation X-ray instruments, such as HEX-P, AXIS, or Strobe-X, should be able to reach weaker short bursts in the nearby universe, constraining further the magnetar nature of nearby FRBs, including FRB 20200120E \citep{Alford24FrASS}.

\subsection{Magnetar Giant Flare Spikes}
\label{sec:mgfspikes}

In the 1970s the debate on the origin of gamma-ray bursts (GRBs) was an outstanding question in astrophysics. Key pieces of information were that none were associated to known sources, none had been shown to repeat,  and that their lightcurves seemed to have spiky but random behavior. The arrival of GRB\,790305B was the first GRB localized to a known position (a supernova remnant in the Large Magellanic Cloud), was followed by a weaker GRB from the same position, and an incredibly bright spike was seen followed by a periodic, exponentially decaying tail \citep{mazets1979observations}. The tail period is approximately 8 seconds, complemented by a weaker interpulse occurring at a phase of 0.5 \citep{mazets1979observations, cline1980detection}. The rapid rise time of less than 0.25 ms was the fastest ever seen \citep{cline1980detection}. We now know this to be the first Magnetar Giant Flare (MGF) seen, and the first signal from a magnetar identified. Since then two more flares have been identified from magnetars in the Milky Way \citep{hurley1999giant,palmer2005giant}. All three show similar characteristics, with tails lasting for hundreds of seconds. Due to their extreme luminosities, the spikes of these three giant flares saturated all viewing detectors.

MGFs are the most luminous transients created by magnetars. The crust of the magnetar may store significant elastic energy, which is released when the crust, stressed and powered by the internal magnetic field energy density, deeply and widely fractures \citep{2015MNRAS.449.2047L}. Magnetic reconnection may occur in the magnetosphere releasing a bright spike where the plasma blows off on open field lines, followed by a periodic tail caused by the emission form a plasma fireball magnetically trapped on the rotating surface of the magnetar \citep{Paczynski1992, duncan1992formation}.

By building and characterizing a larger population of MGFs it will be possible to place better constraints on their intrinsic rates, energetics distribution, and maximal energy release. The rates are of key importance to understand the possibility of detection via GWs during future observing runs \citep{LIGOScientific:2019ccu, Macquet:2021eyn} and the possibility that intermediate or giant flares may produce cosmological FRB \citep{Popov_2018, bochenek2020Nat}. The rates are key to understand the formation channels and the fraction of magnetars that emit giant flares, allowing us to understand the processes which produce the most powerful magnets in the cosmos. The rates and energetics distribution will determine if the giant flares are the extreme events of the same underlying population which produces SGR short bursts, or if they are fundamentally distinct. The maximal energy release can be related to the maximal surface magnetic field of magnetars. 

Further, a sample of events allows for testing of theories on the physical mechanisms which power the prompt spikes. However, Galactic events saturate any reasonable GRB monitor, precluding spectral and temporal properties of the spikes at the brightest intervals.  This saturation has prevented the study of whether giant flares only occur with single pulses, or if they show the same internal pulse variability seen in typical and intermediate SGR short bursts. 

Galactic events likely only occur every few decades. In order to substantially grow the sample during our lifetimes we must look to recover and study extragalactic events. These are also key events to study for spectral and temporal properties as they are often sufficiently far to avoid significant saturation effects in GRB monitors. Given their initial spikes' exceptionally high peak luminosities, instruments with high sensitivity, such as the \textit{Fermi} Gamma-ray Burst Monitor \citep[GBM][]{meegan2009fermi} or the Swift Burst Alert Telescope \citep[BAT:][]{2005SSRv..120..143B}), can detect MGF emissions from magnetars located in galaxies possibly up to distances of $25,\rm{Mpc}$ \citep{burns2021identification}. However, the periodic tail ``smoking gun'' signature is not yet recoverable far beyond the Milky Way. Even with more sensitive detectors which may be able to see the tails to the local group, the majority of events they detect will be seen only via their initial spikes. 

Thus, identification of extragalactic giant flares requires reasonably precise localizations and comparison with nearby galaxy catalogs. Six candidate events at differing degree of significance events have now been found: GRB\,070201 from M31, GRB\,051103 and\,231115A from M82, GRB\,070222 from M83, and GRB\,180128A and GRB\,200415A from NGC\,253 \citep{ofek2007soft,2007AstL...33...19F, mazets2008giant, 2008ApJ...681.1464O, hurley2010new, svinkin2021bright, 2021Natur.589..207R, burns2021identification,trigg2023grb}. This spatial alignment method and expectation of extragalactic MGFs masquerading as cosmological short GRBs dates back decades \citep{Hurley2005Natur.434.1098H}. GRB\,051103 and GRB\,070201 were the only candidates identified prior to 2020. Population analyses considering localizations of all short GRBs by Swift and the InterPlanetary Network (IPN) against galaxy catalogs failed to identify additional candidates. The discovery of GRB\,200415A led to the development of an improved search method, weighting possible host galaxies by star formation rate and distance based on the brightness of the GRB, which identified GRB\,070222 in archival data \citep{burns2021identification}. Additionally applying selections to short GRBs including the rise time and duration, both preferentially shorter for MGFs, identified GRB\,180128A \citep{trigg2023grb}. Recently, INTEGRAL detected, localized to few arc minutes ---which is orders of magnitude better than the second best-localized MGF--- enabling rapid follow-up observations, and promptly identified GRB\,231115A as a MGF \citep{mereghetti2023magnetar,2023arXiv231214833Y}, which is the first giant flare with rapid follow-up observations. Further analysis of this event is on-going. It is important to notice how, in this case, even in absence of a pulsating tail detection, it was possible to unambiguously identify the origin of the event as a MGF thanks to the precise localization. A well constrained association to a nearby galaxy, in fact, allows for accurate estimate of distances and hence intrinsic energetic of the burst, effectively excluding other typically more energetic progenitors. 

Constructing a population of MGFs is key for several reasons. Study of Galactic and extragalactic MGFs allows for more precise measures on rates and intrinsic energetics functions \citep{burns2021identification}, which will tell us if these giant flares are the extreme end of the SGR short burst distribution or fully distinct. These measures are also key to understand if MGFs can power FRBs. The study of individual events provides precise temporal and spatial information for deep multimessenger searches. Recovery of the MGF signal, individually or stochastically \citep{Macquet:2021eyn,kouvatsos2022detectability}, allows measure of the f-mode frequency, giving insight into the structure of NSs and their equation of state \citep{kunjipurayil2022impact}. The study of extragalactic MGFs allows for careful (unsaturated) study of their temporal and spectral evolution, allowing insight into their physical origin \citep{trigg2023grb}. Lastly, identifying extragalactic MGFs is the easiest, possibly only way to study magnetars beyond the Magellanic Clouds.

All of these scientific results support the need for continuous, sensitive, all-sky monitoring of the gamma-ray sky. Reasonable localization accuracy is necessary to enable follow-up searches across and beyond the electromagnetic spectrum. The possible harder spectrum of brighter bursts may be key to driving sensitivity at higher energies than typical GRB monitors. Coverage of gamma-rays above the MeV regime are needed to search for more GeV flares, similar to the one found after GRB\,200415A \citep{LATMGF}, which may inform or reject the bow-shock origin that was put forward in \citep{LATMGF}.

\subsection{Pulsating MGF tail from extragalactic magnetars}

The initial spike of the three confirmed MGFs was closely followed by a bright ($L_{\rm X}\approx10^{43}$~erg~s$^{-1}$) thermally emitting ($kT\approx10$~keV) tail, declining quasi-exponentially below the sensitivity of large field-of-view hard X-ray monitors in about 300~seconds). The rotational motion of the NS induces periodic modulation to this tail at the star spin period, providing the smoking-gun evidence for the magnetar central engine of these extreme events. 

These tails are thought to be generated due to the magnetosphere of the NS trapping a fraction of the energy released by the initial burst (likely when magnetic pressure overcomes the radiation pressure as emission from the initial spike abates). This trapped fireball of photon-pair plasma is optically thick and slowly releases energy from its surface as it cools and shrinks in size \cite[essentially evaporating, e.g.,][]{thompson96ApJ}. Observationally, the tail spectra in the 1-100 keV range are dominated by a thermal component with observed temperatures of the order of tens of keV, which decreases with time. A non-thermal component is also present, most prominently at early times and dominating the emission at higher energies ($\gtrsim100$~keV, Boggs et al. 2007).

The spectra of the time-integrated tails of the three MGF tails was compatible with a dominant blackbody component ($k$T of tens keV) and a subdominant powerlaw only emerging above 30--40 keV \citep[see, e.g., figure 4 of][]{hurley2005exceptionally}. The intrinsic total radiative energy of the three observed MGF tails hovers around a few $10^{44}~\mathrm{erg}$ \citep{2008AandARv..15..225M}, despite the fact that the energy from their initial spikes varies by two orders of magnitude. This raises the intriguing question of whether MGF tails are standard candles. The current statistics of the available observations limits our capability to provide a meaningful answer. However the relevance of this realization has important implications for both cosmology (providing a tool for more accurate distance measurements), and the measurement of the, largely unknown, magnetic Eddington limit \citep{2015RPPh...78k6901T}. Additionally, quasi-periodic oscillations (QPOs) at several differing frequencies have been discovered in the tail emission of the Galactic MGFs of SGR 1900+14 and SGR~1806$-$20 \citep[e.g.,][]{2005ApJ...628L..53I, 2005ApJ...632L.111S}. If interpreted as oscillation modes in the NSs crusts, these QPOs could be utilized to place limits on the dense matter equation of state, complimenting other major efforts such as the light curve modeling of millisecond pulsars by NICER \citep{miller19ApJ, riley19ApJ} and the waveform modeling of the gravitational wave signal from double NS mergers \citep{2017PhRvLnsm}. %\textcolro{red}{(providing clues on the role of the magnetosphere in trapping a hot plasma ball)}. 
Thus, expanding our detection capability for MGF tails beyond our galaxy and immediate neighborhood will substantially increase the sample size of these events, in turn providing crucial data to test these tails as an independent cosmological probe, and infer the Eddington limit of highly-magnetized NS.

\begin{figure*}[t]
    \centering
    \includegraphics[width=0.45\textwidth]{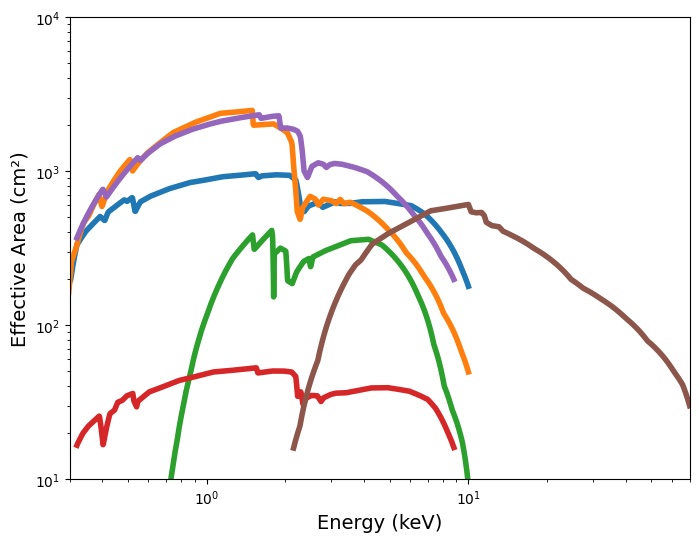}
    \includegraphics[width=0.41\textwidth]{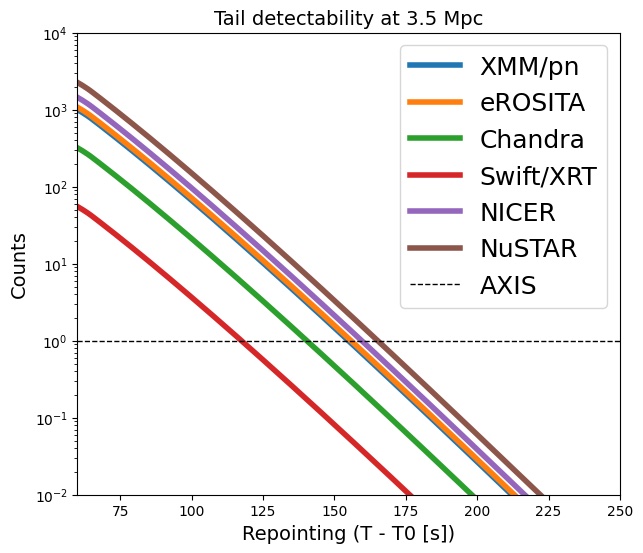}
    \caption{Right: Expected number of counts for time-integrated observation of a MGF tale as it would appear at a distance of 3.5 Mpc (e.g., NGC 243) as a function of repointing time after the initial MGF spike.  We assume a MGF tail similar to the one observed in 1998 from SGR~1900+14 \cite{2001ApJ...549.1021F}, scaled to 3.5 Mpc. Left: Effective areas of different instuments considered are shown for comparison.}
    \label{fig:tails}
\end{figure*}

To this end, we simulate the possible detection of MGF tails with currently operating X-ray satellites, scaled to the extragalactic distance of 3.5 Mpc (e.g., the distance of the star forming galaxies M82 and NGC 253). We assume an event like the 1998 MGF from SGR~1900$+$14 as presented in \cite{2001ApJ...549.1021F}, in which the spectrum is modelled as a blackbody with temperature decreasing over time. We use the effective areas of the instruments as presented in the left panel of Figure~\ref{fig:tails}. The right panel of Figure~\ref{fig:tails} displays the number of expected signal counts as a function of a hypothetical repointing time starting at 60 seconds and integrating over the duration of the tail (~300 s). In this time window, for most pointed X-ray telescopes, the expected background counts is of the order of a few (not included in our simple calculations, as detailed simulations are reserved for an upcoming publication). At 3.5~Mpc, all instruments are capable of detecting the tail assuming a relatively fast repointing, e.g., that of XRT aboard Swift. A NuSTAR or NICER-like instrument, under the same circumstances, could provide a detection up to about 35 Mpc. With an MeV sensitive mission which could detect MGF spikes up to these distances and beyond, an X-ray follow-up instrument with the above capabilities could provide smoking-gun evidence for a population-size sample of MGF paving the way for a major leap towards understanding of these phenomena. On the other hand, a large field-of-view X-ray instrument such as eROSITA \citep{2021A&A...647A...1P} or one equipped with a sensitive lobster eye optic, such as Einstein Probe \citep{2022hxga.book...86Y}, might be able to detect MGF tails independently, and providing an estimate of ``orphan'' MGF tails where the spike emission is beamed away from the observer.

\subsection{Polarization of magnetars' bursts and flares}
X-ray polarization of magnetars traces the magnetic field geometry as well as the the shape, dimension and physical state of the surface emitting region, and exotic effects of quantum electrodynamics (QED) that are expected to take place in presence of extreme magnetic fields like those of magnetars. Despite the great recent observational advancements with the NASA's IXPE mission \citep{2022JATIS...8b6002W}, the highly degenerate parameters space prevents from definitive conclusions on QED effects \citep[see, e.g.,][]{2022Sci...378..646T, 2023ApJ...944L..27Z}. IXPE results, which focus primarily on persistent emission, highlight the need for further theoretical effort and advancements in numerical simulations to build more accurate models. Furthermore, the impossibility in decoupling QED effects on polarization from the geometrical polarization expected when the emitting region is a small patch on the magnetar's surface, seems to suggest the need for extending the range of measured polarization below 1 keV. Probing the polarization from the cooler X-ray radiation emitted from a wider portion of the surface is expected to be a better probe of QED effects. In this context, the further advancement of technologies like the ones developed for REDSoX \citep{2023HEAD...2010367M} ---the Rocket Experiment Demonstration of a Soft X-ray Polarimeter, recently approved by NASA--- will be critical in this endeavour.

Extant models of magnetar burst polarizations are sparse \citep{2017MNRAS.469.3610T} and are currently in development (see, e.g., \cite{2023HEAD...2011649W}). The combination of different outgoing photon angles sampled by the observer on a magnetic loop, however, is expected to reduce the time-averaged polarization of the bursts to around $30-60\%$. Polarization of bursts is also expected to be energy-, viewing- and magnetic-geometry dependent with possible influences of gravitational lensing by the magnetar. Any actual observational constraints on burst polarization, combined with measured broadband spectra of high-energy monitors can greatly inform the factors influencing burst polarization such as magnetar viewing geometry, size of the active flux tube, and rotational spin phase of the burst. This, in turn, combined with other high-energy observations, can elucidate the active region and physics of the magnetar crust.  As magnetar bursts are sporadic and unpredictable, catching a bright burst serendipitously in pointed observations is unlikely, unless a burst storm is ongoing, in which case assessment of the polarization could be limited by counts statistics for individual short bursts, or pile-up in case of extremely bright events. A way to observe this with current instruments (such as IXPE) is through the delayed X-ray emission from the very bright bursts scattered off dust layers along the line of sight. Such observations would be effective in the approximation that the dust-scattering-induced polarization is negligible ($sim10^-5$ modulation), which is valid for small (arcminutes) scattering angles \citep[see appendix B.2 of][]{negro2023ixpe}. A fast-pointing soft X-ray polarimeter or a sensitive monitor with polarization capability would greatly widen our observational portfolio, allowing for more modeling and better understanding of the processes involved in Galactic magnetars burst activity.

In the context of MGFs, polarization observations would provide the key information about structure of the magnetars magnetosphere. Ref.~\citep{2017MNRAS.469.3610T} models the spectral and polarization properties of the 1-100 keV radiation emitted during the MGF tails invoking a simplified ``trapped-fireball'' model, in which the electron-positron pair plasma is injected in the magnetosphere and remains trapped within the closed lines of the strong magnetic field. The linear polarization predicted by this model is very high (greater then 80\% between 1-100 keV). Ref.~\cite{2017MNRAS.469.3610T} adopts a similar model to predict the linear polarization from  MGF tails, assuming a more realistic temperature distribution in the fireball but integrating over wider energy ranges, finding a lower polarization degree, as high as 30\% (1--30 keV) and 10\% (30--100 keV) depending on the viewing angle with respect to the magnetic axis of the magnetar.

% assume magnetic scattering as the main source of opacity in the fireball medium; 
% Among the additional second-order processes that can take place in magnetized plasma there are thermal bremsstrahlung, photon splitting and double-Compton scattering; The bremsstrahlung between particles with the same charge becomes important only for high-energy particles (above 300 keV; only the electron-positron bremsstrahlung, slightly enhanced with respect to the electron-ion one [23,24], turns out to be relevant (but still ignored because the amplitude of the process is negligible). The model reproduces the spectral features and the periodic behaviour of the MGF, however that is the case for many models: the discrimination power comes down to the polarization (see Zane et al. 2023).
% \begin{figure*}
%     \centering
%     \includegraphics[width=0.5\textwidth]{trapped_fireball.jpg}
%     \caption{Figure from Ref.~\cite{2017MNRAS.469.3610T} showing the structure of the trapped fireball. The dark orange region represents the inner part of the trapped fireball, while the atmosphere of the trapped fireball, the region from which most of the observed photons are coming from, is illustrated lighter shade of orange.}
%     \label{fig:fireball}
% \end{figure*}
Such discrepancy of different predictions highlights how polarization measurements of MGF tails could help constrain the trapped-fireball model and potentially drive new theories to explain magnetar flares. Such observations are not possible in the soft X-ray band as IXPE could not repoint fast enough to catch the emission, while at higher energies, where the future missions COSI and POLAR 2 will operate, theoretical predictions are lacking. COSI---the COmpton Spectrometer and Imager \citep{2023arXiv230812362T}--- is scheduled to launch in 2027, will be sensitive to soft gamma rays between 200 keV and 5 MeV, and will have polarization capabilities for assessing Galactic MGF tails. A dedicated study on COSI's detection capabilities of extragalactic MGFs is needed.

\section{Magnetars in Multimessenger Astronomy}
\label{sec:3}
\subsection{Gravitational waves from magnetar bursts and flares}
Gravitational waves (GWs) from magnetars can be generated through various astrophysical processes that involve rapid changes in the mass distribution or extreme deformations of these highly magnetized NSs. The intense magnetic fields associated with magnetars, significantly influence their dynamics and can give rise to GWs emissions. This is believed to happen when the intense magnetic fields of magnetars undergo instabilities, causing dramatic reconfigurations. The associated GW waveforms depend on the specifics of the starquake and the characteristic frequencies are unknown. MGFs are thought to excite two different types of oscillations, the fundamental (of f-mode), which radiate GWs, and the shear modes or torsional modes, that manifest themselves with observable QPOs. The f-mode is thought to be excited when the magnetars internal magnetic field rearranges itself, while QPOs are other excited oscillation modes most likely due to seismic vibrations, and are longer-lived than the f-mode. QPOs have been detected in the tail emission of all three nearby MGFs \citep{2005ApJ...628L..53I, 2005ApJ...632L.111S, 2006ApJ...637L.117W, 2006ApJ...653..593S}, and, interestingly, QPOs in short repeated bursts from SGR~J1550-5418 were also reported in 2014 by \cite{2014ApJ...787..128H}. In general, however, the frequencies detected are disparate and the vibration modes are difficult to identify given the numerous stellar parameters involved (magnetic field, mass, radius, composition, etc...) and the rarity of these events.

While GWs are generally anticipated to accompany energetic bursts, this expectation is especially pronounced and accessible in the case of MGFs, representing the most intense starquakes in magnetars.  This expectation is predicated upon the assumption that mass redistribution can yield a GW luminosity that is a sizable fraction of the total radiative luminosity of $10^{45}-10^{47}\,$erg/sec in the initial spike. Such GW luminosities are readily accessible to LVK for magnetars in the Milky Way and in nearby galaxies.  Despite these expectations, the detection of GWs from MGFs remains elusive, with none having been observed to date \citep[see, e.g.,][for the search in the previous LIGO-VIRGO-KAGRA observing run]{O3LVKmagnetar}. In 2004, in occasion of the MGF from SGR~1806--20 \citep{palmer2005giant}, the early LIGO interferemeters reported only upper limits \citep{abbott2007search} on a possible GW emission. As pointed out in the Gamma-ray Transient Network Science Analysis Group \citep{2023arXiv230804485B}, the current GW detector network is about two orders of magnitude more sensitive than the first generation, and another factor of 100 is expected within the next 20 years of upgrades. Such improvement from the GW front can lead the first detection of GWs from magnetars in our Galaxy and beyond. In this context, the presence of wide field of view high-energy monitors with a fast turnaround is imperative  to promptly detect electromagnetic counterparts. Such observations would constrain the total energy that can be radiated via GW, as well as the ratio between electromagnetic energy vs GW energy during magnetar flares, providing major advances in our understanding of magnetars (and NSs in general), constraining the models of matter structure and behaviours in such extreme environments.

GWs are also likely produced at the magnetar's birth.  In Section \ref{sec:mgfspikes}, we touched upon the relevance of observing a second MGF from the same magnetar, in terms of being the first source of repeating GRBs. However, another implication of repeating MGFs, as pointed out by Ref.~\cite{2005ApJ...634L.165S}, is the requirement of a magnetic field above $10^{16}$ G of newly born magnetars. Such extreme internal field necessarily deforms the NS; if its moment of inertia has axes not aligned with the rotational axis, it would generate a week-long strong gravitational wave signal. The frequency of such a GW signal is dictated by the fast rotation period of the new-born magnetar. Ref.~\cite{2005ApJ...634L.165S} predicted the detection of such GW signal by Advance LIGO-class detectors up to the distance of the Virgo Cluster ($\sim$ 2000 galaxies), where magnetars are expected to form with a rate of more than one magnetar per year. GW detections of newborn magnetars \citep[see also][]{2020MNRAS.494.4838L}, have so far not been forthcoming.

Models predicting gravitational wave signals from magnetars \citep[see, e.g., ][and references therein]{2012ApJ...760....1C, 2022ASSL..465..245D} face considerable uncertainty due to our limited understanding of their internal magnetic field configurations and matter equations of state. This uncertainty spans from optimistic to pessimistic expectations. Further investigation into magnetars' transient activity holds promise in elucidating the underlying physics, potentially improving prediction reliability.
% \textcolor{red}{can we say something about the implications of this "non observation"?}

\subsection{Neutrinos from magnetars}

During the initial phases of a magnetar flare or burst, the intense release of energy can heat the NS's crust and interior. Subsequent cooling processes, involving neutrino emission, become prominent. Neutrinos, being weakly interacting particles, can escape the dense magnetar environment and carry away significant amounts of energy. We can distinguish between high-energy neutrinos, of GeV--TeV energy, detectable by instruments like the IceCube Observatory \citep{2006APh....26..155I}, and MeV neutrinos, like the ones produced in stellar processes and supernovae explosions, detectable by instruments like Super-Kamiokande \citep{2008nops.book...19W}. Both classes of neutrinos, when detected in coincidence with the electromagnetic counterpart, are a crucial aspect of multimessenger astronomy ---which, in a sense, can be dated back to the detection of MeV neutrinos from SN 1987A \citep{1987ApJ...320..589B}. In the context of magnetar bursts and flares, models have been developed to predict the emission of high-energy neutrinos, whose detection would provide important information about the flaring mechanism, as well as the crustal composition. In general, the production of neutrinos requires the presence of hadronic or photo-hadronic interactions. In MGFs the neutrino fluxes depend on the baryon load, which is not well constrained, due to uncertainties on the relative importance of thermal and non-thermal components \citep{2005ApJ...633.1013I}. Hence, detection of neutrinos from magnetars would be extremely insightful to understand their composition.

\begin{figure*}[t]
    \centering
    \includegraphics[width=0.48\textwidth]{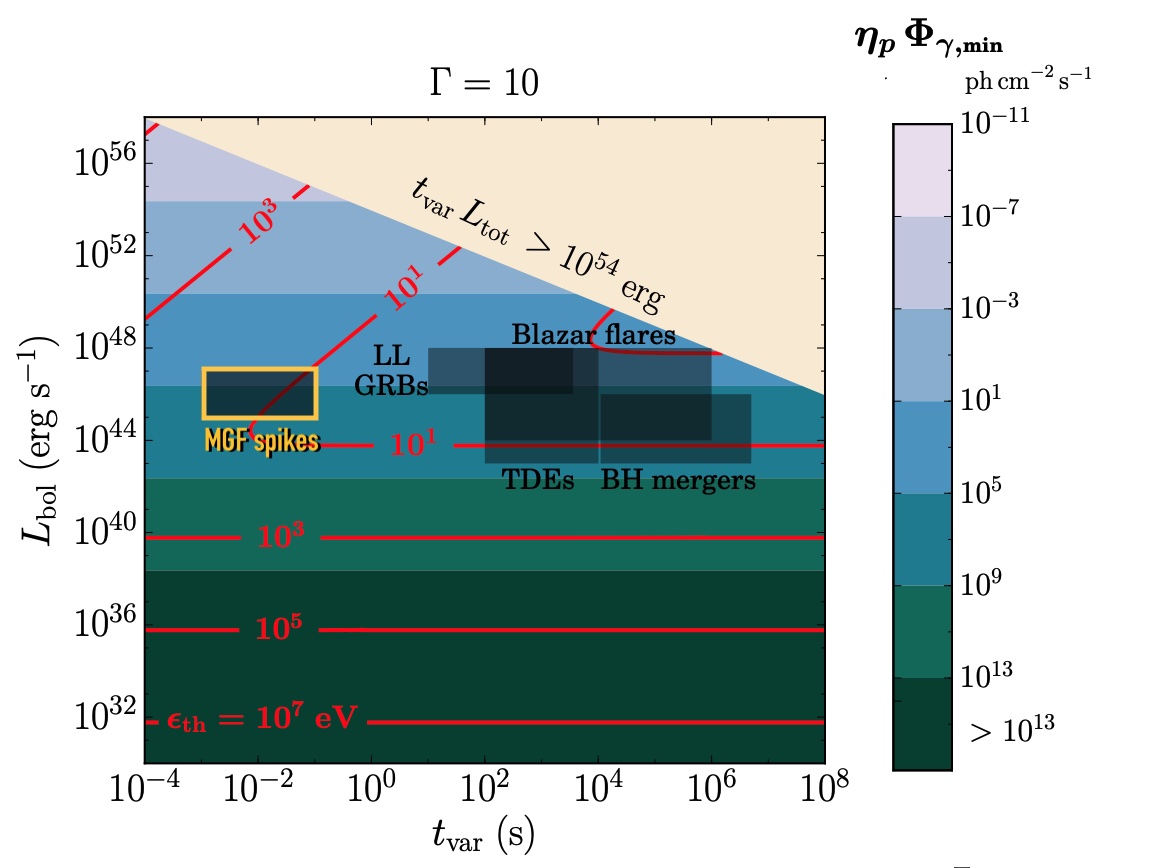}
    \includegraphics[width=0.48\textwidth]{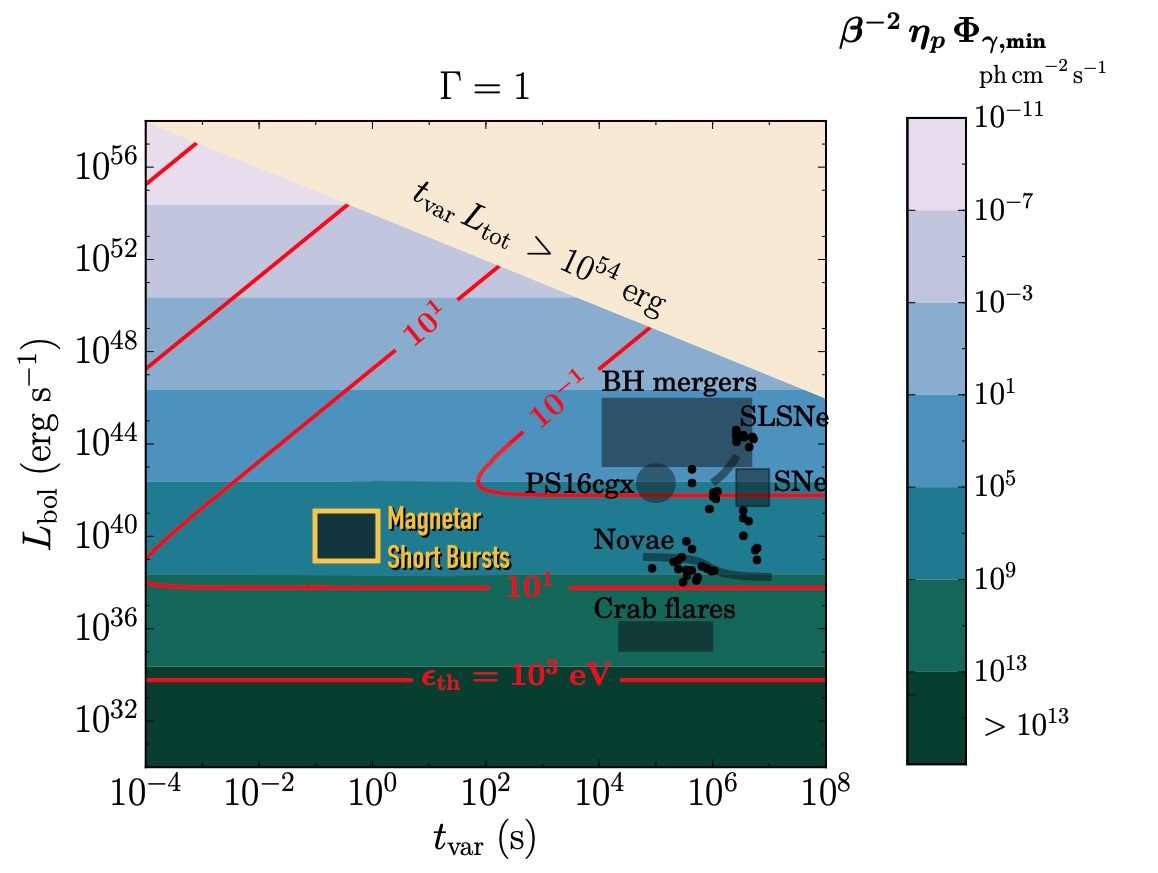}\\
    \caption{Adapted from \cite{2017A&A...603A..76G} (Fig.~1). The two plots illustrate the minimum photon flux needed to have a neutrino detection in IceCube detectors for different Lorentz factors. Highlighted with yellow boxes are the magnetars' transients considered, short bursts and MGF spikes.}
    \label{fig:Guepin}
\end{figure*}

% High-energy neutrinos from magnetars bursts and flares can be generated via photo-hadronic interaction---protons from the star interact with thermal photospheric photons producing pions, which then decay producing neutrinos) \citep[see, e.g.,]{2005ApJ...633.1013I}. 
% Ref.~estimated neutrino fluxes from the 2004 MGFs from SGR~1806-20 assuming a typical fireball model, considering proton-proton interactions and photo-hadronic interactions with photospheric thermal radiation. They found upper limits for IceCube.
One can build the expectation of the high-energy neutrino yield knowing the expected photon flux of the outflow. This was done in Ref.~\cite{2017A&A...603A..76G}, where they computed the minimum photon flux necessary for neutrino detection by IceCube, as well as the maximum neutrino energy expected, for a number of different sources of outflows (including magnetars bursts and flares). The study is generalized in terms of the intrinsic bolometric luminosity, the Lorentz factor, and the time variability of the emission. In Figure~\ref{fig:Guepin} we highlighting the results for magnetars' transient activity. This study shows how neutrino detection is limited to only very nearby bright events, i.e., MGFs with maximum luminosity distance of $\sim$0.39 Mpc (minimum photon flux of $10^4-10^6$ ph cm$^{-2}$s$^{-1}$ to have a neutrino detection in IceCube). The procedure followed by \cite{2017A&A...603A..76G} is somewhat simplistic and assumes high hadronic yield and maximally efficient proton acceleration associated with relativistic outflows. This might be attained for MGFs, but it's unlikely for short bursts. Models predict relativistic outflows in the tails of MGFs \citep{2016MNRAS.461..877V} as necessary ingredient to reproduce the observed pulse fraction, offering therefore prospects for high-energy neutrinos emission during the tail-phase of MGFs if proton acceleration is tenable. As pointed out by \cite{2005ApJ...633.1013I}, if TeV neutrinos are detected, one would also expect detectable EeV cosmic rays and possibly TeV gamma-ray emission in coincidence. No claim of such detection has been made so far. 

The work presented in Ref.~\citep{2023arXiv230715375G} searched for high-energy neutrinos from Galactic magnetars, performing a time-integrated search over 14 years of data collected by the IceCube Observatory \citep{IceCube}. The results point out that a next generation upgrade of the neutrino detector with improved sensitivity is in order, as the current IceCube capabilities are $\sim$ two orders of magnitude above the needed sensitivity to detect a stacked signal from all known magnetars. The creation of a magnetar bursts catalog would be beneficial for targeted time-dependent neutrino searches anticipated in \cite{2023arXiv230715375G} as future studies. 

\subsection{Link to Fast Radio Bursts}
\label{sec:magnetar_frb}
FRBs are extragalactic flashes of radio emission of millisecond duration of isotropic-equivalent energies $10^{36}-10^{41}$ erg, first\footnote{Although possibly much earlier by \cite{Linscott1980}.} reported by \cite{lorimer2007bright}. Only recently (since about 2014)  has their true astrophysical nature been accepted, over instrumental backgrounds or artifacts \citep{2014ApJ...790..101S}. FRBs have now become major interest of study and industry in radio astronomy \citep{2021Univ....7..453C}, with propagation effects particularly useful cosmological probes \citep{2014PhRvD..89j7303Z,2018NatCo...9.3833L} such as the baryon fraction of the intergalactic medium \citep{Macquart20}. They are also currently an eminent topic in time-domain astronomy. Many future facilities prominently feature FRBs, or radio transients more broadly, as one of their key science topics. Yet, as it will be clear below, there is a intimate connection with magnetars and their soft gamma-rays short bursts.

Magnetars were initially proposed as the engines of the 2001 Lorimer burst among many models, although in the form of giant flares producing FRBs \citep{2010vaoa.conf..129P,Popov&Postnov13}. Yet various non-magnetar and exotic models were also proposed \cite{Platts+19}. As the first repeating FRB was discovered \citep{2016Natur.531..202S}, giant flares from ``hyperactive magnetars became a popular model \citep[e.g.,][]{2017ApJ...843L..26B,2019MNRAS.485.4091M} over cataclysmic events. However, statistics of waiting times and power-law distributions of fluence in repeating FRBs suggested much more similarity with magnetar short bursts \citep{2019ApJ...879....4W}. Yet, as of 2019, no FRBs were seen from many thousands of short bursts recorded from known magnetars in our local universe. Moreover, radio limits on the SGR 1806-20 giant flare in 2004 ruled out any contemporenous bright radio flashes \citep{2016ApJ...827...59T}; thus it appears giant flares do not necessarily produce FRBs. As suggested by \citep{2019ApJ...879....4W,2020ApJ...891...82W}, special conditions (e.g. charge starvation and pair cascades in the magnetosphere) must be satisfied such that not all short bursts produce radio emission (yet all FRBs would be associated with short bursts, as the FRB occurs in the beginning ``clean" stage of the fireball created in short bursts). As FRBs are result from coherent emission processes, and short bursts are incoherent, the energy contained in FRBs is generally expected to be a small fraction of the total energy observed in the quasi-thermal short bursts. The same conditions thought to be conducive to the production of FRBs (i.e. explosive pair production demanded by large coherent electric fields) likely are also suitable for proton acceleration and the production of high energy neutrinos at low altitudes in the magnetar magnetosphere \citep[e.g.,][]{2008JCAP...08..025H}.

The situation was clarified dramatically in April 2020, when SGR 1935+2154 underwent a burst storm emitting thousands of short bursts in the hard X-rays \citep{2020ATel13675....1P,2020ApJ...904L..21Y}. In the waning hours of this storm, CHIME/FRB \citep{chime20Natur} and STARE2 \citep{bochenek2020Nat} observed a bright radio flash consistent with a FRB from SGR 1935+2154. The radio burst was bright enough (energy $\sim 10^{36}$ erg isotropic equivalent) if placed at cosmological distance to be similar to weaker extragalactic FRBs. Thus, at least a fraction of FRBs originate from magnetars. The burst featured a bright ($10^{40}$ erg) and prompt hard X-ray short burst counterpart detected by INTEGRAL, HXMT-Insight and Konus Wind \citep{ridnaia21NatAs,mereghetti20ApJ,liNatAs1935} (although not by Fermi-GBM and Swift-BAT due to Earth occultation). The HXMT-Insight light curve of the FRB-associated burst featured a $~30-40$ Hz quasi-periodic oscillation \citep{2022ApJ...931...56L} consistent with a low-order crustal torsional eigenmode of a NS, bolstering the case that FRBs are related to  magnetar crustal dynamics and how that is transmitted to the magnetosphere \citep{2020ApJ...903L..38W}. Moreover, the radio led features in the short burst counterpart by a few milliseconds, suggesting a magnetospheric origin to this radio burst, and perhaps all FRBs \citep{2023ApJ...953...67G,2023arXiv231016932G}. More recent statistical ``aftershock" analyses of extragalactic FRBs and SGR 1935+2154 have revealed similarities, to each other, and to earthquake dynamics (but not solar flare catalogs) \citep{2023MNRAS.526.2795T,2024arXiv240116758T} suggesting that the crustal dynamics on magnetars are key to understanding FRBs. 

Radio activity in SGR 1935+2154 is also connected with torque and potentially interior dynamics of the magnetar. In October 2020, SGR 1935+2154 became radio active again, exhibiting bright radio bursts \citep{2021NatAs...5..414K} as well as a prolonged episode of pulsar-like pulsed radio emission \citep{2023SciA....9F6198Z}. This is suggestive of conditions which are conducive to both phenomena, and a magnetospheric origin of FRBs. For this episode, X-ray timing revealed a jump in the period of the magnetar \cite{younes23NatAs} i.e. a spin-down glitch, consistent with a baryon loaded wind extracting angular momentum from the star. More recently, \cite{Hu-2024Natur.626..500H} report  X-ray timing revealing two spin-up glitches separated by $\sim9$ hours bracketing FRB-like radio bursts \citep{2022ATel15681....1D,2022ATel15697....1M} during an epoch of waning burst rate but high spin-down in October 2022. This result suggests have a high superfluid fraction of the magnetar, burst activity possibly triggering the first spin-up glitch. The glitch, in turn possibly triggered the baryonic wind and magnetospheric conditions conducive for radio bursts.

There are many open questions concerning FRBs and the putative magnetar connection: Why do only a small fraction of magnetar short bursts result in an FRB-like emission? Why are some extragalactic FRB sources much more prolific FRB producers than Galactic magnetars? What is the origin of long-timescale periodic activity windows in extragalactic FRBs \citep[e.g.,][]{2020MNRAS.495.3551R,2020Natur.582..351C}, and is this related to the recently reported Galactic long-period magnetar candidates \citep{2022NatAs...6..828C,2022Natur.601..526H,2023Natur.619..487H,2023MNRAS.520.1872B}? Can magnetars involved in NS mergers produce radio bursts \citep{2023MNRAS.519.3923C}?
To answer these questions,  further study of local magnetars, and extragalactic magnetar signals correlated in time and sky location in multiple messengers will likely be crucial.

\section{Summary and Conclusions}
We conclude by underscoring the critical role of continuous monitoring and real-time detection and alert capabilities in advancing our understanding of magnetars, as well as other transient events in the high-energy astrophysical landscape. 

Long-term monitoring campaigns of magnetars outbursts, particularly in X-rays alongside radio and infrared observations, have yielded invaluable insights into the behavior of magnetars. Swift and NICER have played pivotal roles in this regard, with their continued operation being paramount for future discoveries. Sensitive, continuous monitoring of the high-energy sky plays a crucial role in detecting bursts and flares from magnetars, both alone and in concert with FRB monitoring and possibly future GW and neutrino observations. Increased sensitivity of all-sky monitors could reveal MGF tail emission of extragalactic events, providing the unambiguous signature for a magnetar origin. At the same time, improved localizations could unambiguously exclude a cosmological origin of the detected gamma-ray burst \citep{mereghetti2023magnetar}. Precise localizations may also allow for the determination of repeat giant flares from individual magnetars in other galaxies, a question which has not been resolved directly in 50 years of monitoring the Milky Way. Capturing orphan MGF tail detection, where the spike emission is directed away from the observer, also requires ultra-fast repointing instruments or large field-of-view X-ray instruments equipped with sensitive optics. High-energy polarimetry offers a unique window into the physical processes driving magnetar transients, shedding light on magnetic field configurations, emission mechanisms, and the nature of the emitting sources. The recent non-selection of LEAP ---A LargE Area burst Polarimeter--- by NASA represents a missed opportunity to gather new insights from fast-transient polarization in the 50--500 keV energy range. In general, a wide-field polarimeter with sensitivity down to tens of keV would greatly contribute to enhancing our understanding of magnetars dynamics through the observations of nearby extragalactic MGFs and Galactic intermediate flares.

The aging status of current instruments, including Konus, Swift, and Fermi, coupled with the decomissioning of AGILE and the impending decommissioning of INTEGRAL, underscores the urgency of advancing technology and developing new missions to ensure uninterrupted coverage and enhanced capabilities for detecting fast transients. As technology evolves, there is optimism for improved sensitivity, localization, and monitoring capabilities, paving the way for further discoveries in the dynamic field of high-energy astrophysics.

\section*{Acknowledgement}

GY and ZW's contributions are based upon work supported by NASA under award number 80GSFC21M0002. This work has made use of the NASA Astrophysics Data System.

% \printbibliography
\bibliographystyle{plain}
\bibliography{apssamp}% Produces the bibliography via BibTeX.

\begin{thebibliography}{100}

\bibitem{abbott2007search}
B~Abbott, R~Abbott, R~Adhikari, J~Agresti, P~Ajith, B~Allen, R~Amin, SB~Anderson, WG~Anderson, M~Arain, et~al.
\newblock Search for gravitational wave radiation associated with the pulsating tail of the sgr 1806- 20 hyperflare of 27 december 2004 using ligo.
\newblock {\em Physical Review D}, 76(6):062003, 2007.

\bibitem{LIGOScientific:2019ccu}
B.~P. Abbott et~al.
\newblock {Search for Transient Gravitational-wave Signals Associated with Magnetar Bursts during Advanced LIGO\textquoteright{}s Second Observing Run}.
\newblock {\em Astrophys. J.}, 874(2):163, 2019.

\bibitem{Alford24FrASS}
J.~A.~J. {Alford}, G.~A. {Younes}, Z.~{Wadiasingh}, M.~{Abdelmaguid}, H.~{An}, M.~{Bachetti}, M.~G. {Baring}, A.~{Beloborodov}, A.~Y. {Chen}, T.~{Enoto}, J.~A. {Garc{\'\i}a}, J.~D. {Gelfand}, E.~V. {Gotthelf}, A.~K. {Harding}, C.~P. {Hu}, A.~D. {Jaodand}, V.~{Kaspi}, C.~{Kim}, C.~{Kouveliotou}, L.~{Kuiper}, K.~{Mori}, M.~{Nynka}, J.~{Park}, D.~{Stern}, J.~{Valverde}, and D.~J. {Walton}.
\newblock {The high energy X-ray probe (HEX-P): magnetars and other isolated neutron stars}.
\newblock {\em Frontiers in Astronomy and Space Sciences}, 10:1294449, January 2024.

\bibitem{Archibald2013}
R.~F. {Archibald}, V.~M. {Kaspi}, C.~Y. {Ng}, K.~N. {Gourgouliatos}, D.~{Tsang}, P.~{Scholz}, A.~P. {Beardmore}, N.~{Gehrels}, and J.~A. {Kennea}.
\newblock {An anti-glitch in a magnetar}.
\newblock {\em \nat}, 497(7451):591--593, May 2013.

\bibitem{archibald2016ApJ}
R.~F. {Archibald}, V.~M. {Kaspi}, S.~P. {Tendulkar}, and P.~{Scholz}.
\newblock {A Magnetar-like Outburst from a High-B Radio Pulsar}.
\newblock {\em \apjl}, 829(1):L21, September 2016.

\bibitem{Archibald2020}
R.~F. {Archibald}, P.~{Scholz}, V.~M. {Kaspi}, S.~P. {Tendulkar}, and A.~P. {Beardmore}.
\newblock {Two New Outbursts and Transient Hard X-Rays from 1E 1048.1-5937}.
\newblock {\em \apj}, 889(2):160, February 2020.

\bibitem{atteia87ApJ}
J.~L. {Atteia}, M.~{Boer}, K.~{Hurley}, M.~{Niel}, G.~{Vedrenne}, E.~E. {Fenimore}, R.~W. {Klebesadel}, J.~G. {Laros}, A.~V. {Kuznetsov}, R.~A. {Sunyaev}, O.~V. {Terekhov}, C.~{Kouveliotou}, T.~{Cline}, B.~{Dennis}, U.~{Desai}, and L.~{Orwig}.
\newblock {Localization, Time Histories, and Energy Spectra of a New Type of Recurrent High-Energy Transient Source}.
\newblock {\em \apjl}, 320:L105, September 1987.

\bibitem{2022A&A...668A..79B}
P.~{Barr{\`e}re}, J.~{Guilet}, A.~{Reboul-Salze}, R.~{Raynaud}, and H.~T. {Janka}.
\newblock {A new scenario for magnetar formation: Tayler-Spruit dynamo in a proto-neutron star spun up by fallback}.
\newblock {\em \aap}, 668:A79, December 2022.

\bibitem{2005SSRv..120..143B}
Scott~D. {Barthelmy}, Louis~M. {Barbier}, Jay~R. {Cummings}, Ed~E. {Fenimore}, Neil {Gehrels}, Derek {Hullinger}, Hans~A. {Krimm}, Craig~B. {Markwardt}, David~M. {Palmer}, Ann {Parsons}, Goro {Sato}, Masaya {Suzuki}, Tadayuki {Takahashi}, Makota {Tashiro}, and Jack {Tueller}.
\newblock {The Burst Alert Telescope (BAT) on the SWIFT Midex Mission}.
\newblock {\em \ssr}, 120(3-4):143--164, October 2005.

\bibitem{beloborodov09ApJ}
Andrei~M. {Beloborodov}.
\newblock {Untwisting Magnetospheres of Neutron Stars}.
\newblock {\em \apj}, 703(1):1044--1060, September 2009.

\bibitem{2017ApJ...843L..26B}
Andrei~M. {Beloborodov}.
\newblock {A Flaring Magnetar in FRB 121102?}
\newblock {\em \apjl}, 843(2):L26, July 2017.

\bibitem{beloborodov16ApJ}
Andrei~M. {Beloborodov} and Xinyu {Li}.
\newblock {Magnetar Heating}.
\newblock {\em \apj}, 833(2):261, December 2016.

\bibitem{2023MNRAS.520.1872B}
P.~{Beniamini}, Z.~{Wadiasingh}, J.~{Hare}, K.~M. {Rajwade}, G.~{Younes}, and A.~J. {van der Horst}.
\newblock {Evidence for an abundant old population of Galactic ultra-long period magnetars and implications for fast radio bursts}.
\newblock {\em \mnras}, 520(2):1872--1894, April 2023.

\bibitem{Beniamini2019}
Paz {Beniamini}, Kenta {Hotokezaka}, Alexander {van der Horst}, and Chryssa {Kouveliotou}.
\newblock {Formation rates and evolution histories of magnetars}.
\newblock {\em \mnras}, 487(1):1426--1438, July 2019.

\bibitem{2020MNRAS.496.3390B}
Paz {Beniamini}, Zorawar {Wadiasingh}, and Brian~D. {Metzger}.
\newblock {Periodicity in recurrent fast radio bursts and the origin of ultralong period magnetars}.
\newblock {\em \mnras}, 496(3):3390--3401, August 2020.

\bibitem{bhawdwaj21ApJ}
M.~{Bhardwaj}, B.~M. {Gaensler}, V.~M. {Kaspi}, T.~L. {Landecker}, R.~{Mckinven}, D.~{Michilli}, Z.~{Pleunis}, S.~P. {Tendulkar}, B.~C. {Andersen}, P.~J. {Boyle}, T.~{Cassanelli}, P.~{Chawla}, A.~{Cook}, M.~{Dobbs}, E.~{Fonseca}, J.~{Kaczmarek}, C.~{Leung}, K.~{Masui}, M.~{Mnchmeyer}, C.~{Ng}, M.~{Rafiei-Ravandi}, P.~{Scholz}, K.~{Shin}, K.~M. {Smith}, I.~H. {Stairs}, and A.~V. {Zwaniga}.
\newblock {A Nearby Repeating Fast Radio Burst in the Direction of M81}.
\newblock {\em \apjl}, 910(2):L18, April 2021.

\bibitem{1987ApJ...320..589B}
W.~M. {Blanco}, B.~{Gregory}, M.~{Hamuy}, S.~R. {Heathcote}, M.~M. {Phillips}, N.~B. {Phillips}, N.~B. {Suntzeff}, D.~M. {Terndrup}, A.~R. {Walker}, R.~E. {Williams}, M.~G. {Pastoriza}, T.~{Storchi-Bergmann}, and J.~{Matthews}.
\newblock {Supernova 1987A in the Large Magellanic Cloud: Initial Observations at Cerro Tololo}.
\newblock {\em \apj}, 320:589, September 1987.

\bibitem{bochenek2020Nat}
C.~D. {Bochenek}, V.~{Ravi}, K.~V. {Belov}, G.~{Hallinan}, J.~{Kocz}, S.~R. {Kulkarni}, and D.~L. {McKenna}.
\newblock {A fast radio burst associated with a Galactic magnetar}.
\newblock {\em \nat}, 587(7832):59--62, November 2020.

\bibitem{1997A&AS..122..299B}
G.~{Boella}, R.~C. {Butler}, G.~C. {Perola}, L.~{Piro}, L.~{Scarsi}, and J.~A.~M. {Bleeker}.
\newblock {BeppoSAX, the wide band mission for X-ray astronomy}.
\newblock {\em \aaps}, 122:299--307, April 1997.

\bibitem{2020ApJ...901...18B}
Axel {Brandenburg}.
\newblock {Hall Cascade with Fractional Magnetic Helicity in Neutron Star Crusts}.
\newblock {\em \apj}, 901(1):18, September 2020.

\bibitem{brown09ApJ}
Edward~F. {Brown} and Andrew {Cumming}.
\newblock {Mapping Crustal Heating with the Cooling Light Curves of Quasi-Persistent Transients}.
\newblock {\em \apj}, 698(2):1020--1032, June 2009.

\bibitem{burns2021identification}
E~Burns, D~Svinkin, K~Hurley, Z~Wadiasingh, M~Negro, G~Younes, R~Hamburg, A~Ridnaia, D~Cook, SB~Cenko, et~al.
\newblock Identification of a local sample of gamma-ray bursts consistent with a magnetar giant flare origin.
\newblock {\em The Astrophysical Journal Letters}, 907(2):L28, 2021.

\bibitem{2023arXiv230804485B}
Eric {Burns}, Michael {Coughlin}, Kendall {Ackley}, Igor {Andreoni}, Marie-Anne {Bizouard}, Floor {Broekgaarden}, Nelson~L. {Christensen}, Filippo {D'Ammando}, James {DeLaunay}, Henrike {Fleischhack}, Raymond {Frey}, Chris~L. {Fryer}, Adam {Goldstein}, Bruce {Grossan}, Rachel {Hamburg}, Dieter~H. {Hartmann}, Anna Y.~Q. {Ho}, Eric~J. {Howell}, C.~Michelle {Hui}, Leah {Jenks}, Alyson {Joens}, Stephen {Lesage}, Andrew~J. {Levan}, Amy {Lien}, Athina {Meli}, Michela {Negro}, Tyler {Parsotan}, Oliver~J. {Roberts}, Marcos {Santander}, Jacob~R. {Smith}, Aaron {Tohuvavohu}, John~A. {Tomsick}, Zorawar {Wadiasingh}, Peter {Veres}, Ashley~V. {Villar}, Haocheng {Zhang}, and Sylvia~J. {Zhu}.
\newblock {Gamma-ray Transient Network Science Analysis Group Report}.
\newblock {\em arXiv e-prints}, page arXiv:2308.04485, August 2023.

\bibitem{2022NatAs...6..828C}
Manisha {Caleb}, Ian {Heywood}, Kaustubh {Rajwade}, Mateusz {Malenta}, Benjamin~Willem {Stappers}, Ewan {Barr}, Weiwei {Chen}, Vincent {Morello}, Sotiris {Sanidas}, Jakob {van den Eijnden}, Michael {Kramer}, David {Buckley}, Jaco {Brink}, Sara~Elisa {Motta}, Patrick {Woudt}, Patrick {Weltevrede}, Fabian {Jankowski}, Mayuresh {Surnis}, Sarah {Buchner}, Mechiel~Christiaan {Bezuidenhout}, Laura~Nicole {Driessen}, and Rob {Fender}.
\newblock {Discovery of a radio-emitting neutron star with an ultra-long spin period of 76 s}.
\newblock {\em Nature Astronomy}, 6:828--836, May 2022.

\bibitem{2021Univ....7..453C}
Manisha {Caleb} and Evan {Keane}.
\newblock {A Decade and a Half of Fast Radio Burst Observations}.
\newblock {\em Universe}, 7(11):453, November 2021.

\bibitem{camilo06Natur}
Fernando {Camilo}, Scott~M. {Ransom}, Jules~P. {Halpern}, John {Reynolds}, David~J. {Helfand}, Neil {Zimmerman}, and John {Sarkissian}.
\newblock {Transient pulsed radio emission from a magnetar}.
\newblock {\em \nat}, 442(7105):892--895, August 2006.

\bibitem{1996Natur.382..518C}
Baolian {Cheng}, Richard~I. {Epstein}, Robert~A. {Guyer}, and A.~Cody {Young}.
\newblock {Earthquake-like behaviour of soft {\ensuremath{\gamma}}-ray repeaters}.
\newblock {\em \nat}, 382(6591):518--520, August 1996.

\bibitem{2020Natur.582..351C}
{Chime/Frb Collaboration}, M.~{Amiri}, B.~C. {Andersen}, {Bandura}, et~al.
\newblock {Periodic activity from a fast radio burst source}.
\newblock {\em \nat}, 582(7812):351--355, June 2020.

\bibitem{chime20Natur}
{CHIME/FRB Collaboration}, B.~C. {Andersen}, K.~M. {Bandura}, M.~{Bhardwaj}, et~al.
\newblock {A bright millisecond-duration radio burst from a Galactic magnetar}.
\newblock {\em \nat}, 587(7832):54--58, November 2020.

\bibitem{2012ApJ...760....1C}
Riccardo {Ciolfi} and Luciano {Rezzolla}.
\newblock {Poloidal-field Instability in Magnetized Relativistic Stars}.
\newblock {\em \apj}, 760(1):1, November 2012.

\bibitem{cline1981precise}
T.~L. {Cline}, U.~D. {Desai}, B.~J. {Teegarden}, W.~D. {Evans}, R.~W. {Klebesadel}, J.~G. {Laros}, C.~{Barat}, K.~{Hurley}, M.~{Niel}, G.~{Bedrenne}, I.~V. {Estulin}, V.~G. {Kurt}, G.~A. {Mersov}, V.~M. {Zenchenko}, M.~C. {Weisskopf}, and J.~{Grindlay}.
\newblock {Precise source location of the anomalous 1979 March 5 gamma-ray transient.}
\newblock Technical report, April 1982.

\bibitem{cline1980detection}
TL~Cline, UD~Desai, G~Pizzichini, BJ~Teegarden, WD~Evans, RW~Klebesadel, JG~Laros, K~Hurley, M~Niel, and G~Vedrenne.
\newblock Detection of a fast, intense and unusual gamma-ray transient.
\newblock {\em Astrophysical Journal, Part 2-Letters to the Editor, vol. 237, Apr. 1, 1980, p. L1-L5. Research sponsored by the US Department of Energy and NASA.}, 237:L1--L5, 1980.

\bibitem{collazzi15ApJS}
A.~C. {Collazzi}, C.~{Kouveliotou}, A.~J. {van der Horst}, G.~A. {Younes}, Y.~{Kaneko}, E.~{G{\"o}{\u{g}}{\"u}{\c{s}}}, L.~{Lin}, J.~{Granot}, M.~H. {Finger}, V.~L. {Chaplin}, D.~{Huppenkothen}, A.~L. {Watts}, A.~{von Kienlin}, M.~G. {Baring}, D.~{Gruber}, P.~N. {Bhat}, M.~H. {Gibby}, N.~{Gehrels}, J.~{McEnery}, M.~{van der Klis}, and R.~A.~M.~J. {Wijers}.
\newblock {The Five Year Fermi/GBM Magnetar Burst Catalog}.
\newblock {\em \apjs}, 218(1):11, May 2015.

\bibitem{2023MNRAS.519.3923C}
A.~J. {Cooper}, O.~{Gupta}, Z.~{Wadiasingh}, R.~A.~M.~J. {Wijers}, O.~M. {Boersma}, I.~{Andreoni}, A.~{Rowlinson}, and K.~{Gourdji}.
\newblock {Pulsar revival in neutron star mergers: multimessenger prospects for the discovery of pre-merger coherent radio emission}.
\newblock {\em \mnras}, 519(3):3923--3946, March 2023.

\bibitem{zelati18MNRAS}
Francesco {Coti Zelati}, Nanda {Rea}, Jos{\'e}~A. {Pons}, Sergio {Campana}, and Paolo {Esposito}.
\newblock {Systematic study of magnetar outbursts}.
\newblock {\em \mnras}, 474(1):961--1017, February 2018.

\bibitem{cruces21MNRAS}
M.~{Cruces}, L.~G. {Spitler}, P.~{Scholz}, R.~{Lynch}, A.~{Seymour}, J.~W.~T. {Hessels}, C.~{Gouiff{\'e}s}, G.~H. {Hilmarsson}, M.~{Kramer}, and S.~{Munjal}.
\newblock {Repeating behaviour of FRB 121102: periodicity, waiting times, and energy distribution}.
\newblock {\em \mnras}, 500(1):448--463, January 2021.

\bibitem{2022ASSL..465..245D}
Simone {Dall'Osso} and Luigi {Stella}.
\newblock {Millisecond Magnetars}.
\newblock In Sudip {Bhattacharyya}, Alessandro {Papitto}, and Dipankar {Bhattacharya}, editors, {\em Astrophysics and Space Science Library}, volume 465 of {\em Astrophysics and Space Science Library}, pages 245--280, January 2022.

\bibitem{2020ApJ...903...40D}
Davide {De Grandis}, Roberto {Turolla}, Toby~S. {Wood}, Silvia {Zane}, Roberto {Taverna}, and Konstantinos~N. {Gourgouliatos}.
\newblock {Three-dimensional Modeling of the Magnetothermal Evolution of Neutron Stars: Method and Test Cases}.
\newblock {\em \apj}, 903(1):40, November 2020.

\bibitem{2020ApJ...902L..32D}
C.~{Dehman}, D.~{Vigan{\`o}}, N.~{Rea}, J.~A. {Pons}, R.~{Perna}, and A.~{Garcia-Garcia}.
\newblock {On the Rate of Crustal Failures in Young Magnetars}.
\newblock {\em \apjl}, 902(2):L32, October 2020.

\bibitem{2023MNRAS.523.5198D}
Clara {Dehman}, Daniele {Vigan{\`o}}, Stefano {Ascenzi}, Jose~A. {Pons}, and Nanda {Rea}.
\newblock {3D evolution of neutron star magnetic fields from a realistic core-collapse turbulent topology}.
\newblock {\em \mnras}, 523(4):5198--5206, August 2023.

\bibitem{2023MNRAS.518.1222D}
Clara {Dehman}, Daniele {Vigan{\`o}}, Jos{\'e}~A. {Pons}, and Nanda {Rea}.
\newblock {3D code for MAgneto-Thermal evolution in Isolated Neutron Stars, MATINS: the magnetic field formalism}.
\newblock {\em \mnras}, 518(1):1222--1242, January 2023.

\bibitem{dib14ApJ}
Rim {Dib} and Victoria~M. {Kaspi}.
\newblock {16 yr of RXTE Monitoring of Five Anomalous X-Ray Pulsars}.
\newblock {\em \apj}, 784(1):37, March 2014.

\bibitem{2022ATel15681....1D}
Fengqiu~Adam {Dong} and {Chime/Frb Collaboration}.
\newblock {CHIME/FRB Detection of a Bright Radio Burst from SGR 1935+2154}.
\newblock {\em The Astronomer's Telegram}, 15681:1, October 2022.

\bibitem{duncan1998global}
Robert~C Duncan.
\newblock Global seismic oscillations in soft gamma repeaters.
\newblock {\em The Astrophysical Journal}, 498(1):L45, 1998.

\bibitem{duncan1992formation}
Robert~C Duncan and Christopher Thompson.
\newblock Formation of very strongly magnetized neutron stars-implications for gamma-ray bursts.
\newblock {\em The Astrophysical Journal}, 392:L9--L13, 1992.

\bibitem{2020ApJ...896L..30E}
P.~{Esposito}, N.~{Rea}, A.~{Borghese}, F.~{Coti Zelati}, D.~{Vigan{\`o}}, G.~L. {Israel}, A.~{Tiengo}, A.~{Ridolfi}, A.~{Possenti}, M.~{Burgay}, D.~{G{\"o}tz}, F.~{Pintore}, L.~{Stella}, C.~{Dehman}, M.~{Ronchi}, S.~{Campana}, A.~{Garcia-Garcia}, V.~{Graber}, S.~{Mereghetti}, R.~{Perna}, G.~A. {Rodr{\'\i}guez Castillo}, R.~{Turolla}, and S.~{Zane}.
\newblock {A Very Young Radio-loud Magnetar}.
\newblock {\em \apjl}, 896(2):L30, June 2020.

\bibitem{esposito2021ASSL}
Paolo {Esposito}, Nanda {Rea}, and Gian~Luca {Israel}.
\newblock {Magnetars: A Short Review and Some Sparse Considerations}.
\newblock In Tomaso~M. {Belloni}, Mariano {M{\'e}ndez}, and Chengmin {Zhang}, editors, {\em Timing Neutron Stars: Pulsations, Oscillations and Explosions}, volume 461 of {\em Astrophysics and Space Science Library}, pages 97--142, January 2021.

\bibitem{2001ApJ...549.1021F}
M.~{Feroci}, K.~{Hurley}, R.~C. {Duncan}, and C.~{Thompson}.
\newblock {The Giant Flare of 1998 August 27 from SGR 1900+14. I. An Interpretive Study of BeppoSAX and Ulysses Observations}.
\newblock {\em \apj}, 549(2):1021--1038, March 2001.

\bibitem{2007AstL...33...19F}
D.~D. {Frederiks}, V.~D. {Palshin}, R.~L. {Aptekar}, S.~V. {Golenetskii}, T.~L. {Cline}, and E.~P. {Mazets}.
\newblock {On the possibility of identifying the short hard burst GRB 051103 with a giant flare from a soft gamma repeater in the M81 group of galaxies}.
\newblock {\em Astronomy Letters}, 33(1):19--24, January 2007.

\bibitem{Gavriil+04}
Fotis~P. {Gavriil}, Victoria~M. {Kaspi}, and Peter~M. {Woods}.
\newblock {A Comprehensive Study of the X-Ray Bursts from the Magnetar Candidate 1E 2259+586}.
\newblock {\em \apj}, 607(2):959--969, June 2004.

\bibitem{2023ApJ...953...67G}
M.~Y. {Ge}, C.~Z. {Liu}, S.~N. {Zhang}, F.~J. {Lu}, Z.~{Zhang}, Z.~{Chang}, Y.~L. {Tuo}, X.~B. {Li}, C.~K. {Li}, S.~L. {Xiong}, C.~{Cai}, X.~F. {Li}, R.~{Zhang}, Z.~G. {Dai}, J.~L. {Qu}, L.~M. {Song}, S.~{Zhang}, and L.~J. {Wang}.
\newblock {Reanalysis of the X-Ray-burst-associated FRB 200428 with Insight-HXMT Observations}.
\newblock {\em \apj}, 953(1):67, August 2023.

\bibitem{ge24RAA1935}
Ming-Yu {Ge}, Yuan-Pei {Yang}, Fang-Jun {Lu}, Shi-Qi {Zhou}, Long {Ji}, Shuang-Nan {Zhang}, Bing {Zhang}, Liang {Zhang}, Pei {Wang}, Kejia {Lee}, Weiwei {Zhu}, Jian {Li}, Xian {Hou}, and Qiao-Chu {Li}.
\newblock {Spin Evolution of the Magnetar SGR J1935+2154}.
\newblock {\em Research in Astronomy and Astrophysics}, 24(1):015016, January 2024.

\bibitem{1994ApJS...92..351G}
N.~{Gehrels}, E.~{Chipman}, and D.~{Kniffen}.
\newblock {The Compton Gamma Ray Observatory}.
\newblock {\em \apjs}, 92:351, June 1994.

\bibitem{2023arXiv230715375G}
Ava {Ghadimi} and Marcos {Santander}.
\newblock {Search for high-energy neutrinos from magnetars with IceCube}.
\newblock {\em arXiv e-prints}, page arXiv:2307.15375, July 2023.

\bibitem{2023arXiv231016932G}
Utkarsh {Giri}, Bridget~C. {Andersen}, Pragya {Chawla}, Alice~P. {Curtin}, Emmanuel {Fonseca}, Victoria~M. {Kaspi}, Hsiu-Hsien {Lin}, Kiyoshi~W. {Masui}, Ketan~R. {Sand}, Paul {Scholz}, Thomas~C. {Abbott}, Fengqiu~Adam {Dong}, B.~M. {Gaensler}, Calvin {Leung}, Daniele {Michilli}, Mohit {Bhardwaj}, Moritz {M{\"u}nchmeyer}, Ayush {Pandhi}, Aaron~B. {Pearlman}, Ziggy {Pleunis}, Masoud {Rafiei-Ravandi}, Alex {Reda}, Kaitlyn {Shin}, Kendrick {Smith}, Ingrid~H. {Stairs}, David~C. {Stenning}, and Shriharsh~P. {Tendulkar}.
\newblock {Comprehensive Bayesian analysis of FRB-like bursts from SGR 1935+2154 observed by CHIME/FRB}.
\newblock {\em arXiv e-prints}, page arXiv:2310.16932, October 2023.

\bibitem{2004A&A...417L..45G}
D.~{G{\"o}tz}, S.~{Mereghetti}, I.~F. {Mirabel}, and K.~{Hurley}.
\newblock {Spectral evolution of weak bursts from SGR 1806-20 observed with INTEGRAL}.
\newblock {\em \aap}, 417:L45--L48, April 2004.

\bibitem{gogus16ApJ}
Ersin {G{\"o}{\u{g}}{\"u}{\c{s}}}, Lin {Lin}, Yuki {Kaneko}, Chryssa {Kouveliotou}, Anna~L. {Watts}, Manoneeta {Chakraborty}, M.~Ali {Alpar}, Daniela {Huppenkothen}, Oliver~J. {Roberts}, George {Younes}, and Alexander~J. {van der Horst}.
\newblock {Magnetar-like X-Ray Bursts from a Rotation-powered Pulsar, PSR J1119-6127}.
\newblock {\em \apjl}, 829(2):L25, October 2016.

\bibitem{gourgouliatos2016PNAS}
Konstantinos~N. {Gourgouliatos}, Toby~S. {Wood}, and Rainer {Hollerbach}.
\newblock {Magnetic field evolution in magnetar crusts through three-dimensional simulations}.
\newblock {\em Proceedings of the National Academy of Science}, 113(15):3944--3949, April 2016.

\bibitem{2017A&A...603A..76G}
Claire {Gu{\'e}pin} and Kumiko {Kotera}.
\newblock {Can we observe neutrino flares in coincidence with explosive transients?}
\newblock {\em \aap}, 603:A76, July 2017.

\bibitem{harbel07ApSS}
Frank {Haberl}.
\newblock {The magnificent seven: magnetic fields and surface temperature distributions}.
\newblock {\em \apss}, 308(1-4):181--190, April 2007.

\bibitem{2008JCAP...08..025H}
T.~{Herpay}, S.~{Razzaque}, A.~{Patk{\'o}s}, and P.~{M{\'e}sz{\'a}ros}.
\newblock {High energy neutrinos and photons from curvature pions in magnetars}.
\newblock {\em \jcap}, 2008(8):025, August 2008.

\bibitem{Hu2020}
Chin-Ping {Hu}, Beste {Begi{\c{c}}arslan}, Tolga {G{\"u}ver}, Teruaki {Enoto}, George {Younes}, Takanori {Sakamoto}, Paul~S. {Ray}, Tod~E. {Strohmayer}, Sebastien {Guillot}, Zaven {Arzoumanian}, David~M. {Palmer}, Keith~C. {Gendreau}, C.~{Malacaria}, Zorawar {Wadiasingh}, Gaurava~K. {Jaisawal}, and Walid~A. {Majid}.
\newblock {NICER Observation of the Temporal and Spectral Evolution of Swift J1818.0-1607: A Missing Link between Magnetars and Rotation-powered Pulsars}.
\newblock {\em \apj}, 902(1):1, October 2020.

\bibitem{Hu-2024Natur.626..500H}
Chin-Ping {Hu}, Takuto {Narita}, Teruaki {Enoto}, George {Younes}, Zorawar {Wadiasingh}, Matthew~G. {Baring}, Wynn C.~G. {Ho}, Sebastien {Guillot}, Paul~S. {Ray}, Tolga {G{\"u}ver}, Kaustubh {Rajwade}, Zaven {Arzoumanian}, Chryssa {Kouveliotou}, Alice~K. {Harding}, and Keith~C. {Gendreau}.
\newblock {Rapid spin changes around a magnetar fast radio burst}.
\newblock {\em \nat}, 626(7999):500--504, February 2024.

\bibitem{2014ApJ...787..128H}
D.~{Huppenkothen}, C.~{D'Angelo}, A.~L. {Watts}, L.~{Heil}, M.~{van der Klis}, A.~J. {van der Horst}, C.~{Kouveliotou}, M.~G. {Baring}, E.~{G{\"o}{\u{g}}{\"u}{\c{s}}}, J.~{Granot}, Y.~{Kaneko}, L.~{Lin}, A.~{von Kienlin}, and G.~{Younes}.
\newblock {Quasi-periodic Oscillations in Short Recurring Bursts of the Soft Gamma Repeater J1550-5418}.
\newblock {\em \apj}, 787(2):128, June 2014.

\bibitem{Hurley2005Natur.434.1098H}
K.~{Hurley}, S.~E. {Boggs}, D.~M. {Smith}, R.~C. {Duncan}, R.~{Lin}, A.~{Zoglauer}, S.~{Krucker}, G.~{Hurford}, H.~{Hudson}, C.~{Wigger}, W.~{Hajdas}, C.~{Thompson}, I.~{Mitrofanov}, A.~{Sanin}, W.~{Boynton}, C.~{Fellows}, A.~{von Kienlin}, G.~{Lichti}, A.~{Rau}, and T.~{Cline}.
\newblock {An exceptionally bright flare from SGR 1806-20 and the origins of short-duration {\ensuremath{\gamma}}-ray bursts}.
\newblock {\em \nat}, 434(7037):1098--1103, April 2005.

\bibitem{hurley2005exceptionally}
K~Hurley, SE~Boggs, DM~Smith, RC~Duncan, R~Lin, A~Zoglauer, S~Krucker, G~Hurford, H~Hudson, C~Wigger, et~al.
\newblock An exceptionally bright flare from sgr 1806--20 and the origins of short-duration $\gamma$-ray bursts.
\newblock {\em Nature}, 434(7037):1098--1103, 2005.

\bibitem{hurley1999giant}
K~Hurley, T~Cline, E~Mazets, S~Barthelmy, P~Butterworth, F~Marshall, D~Palmer, R~Aptekar, S~Golenetskii, V~Il'Inskii, et~al.
\newblock A giant periodic flare from the soft $\gamma$-ray repeater sgr1900+ 14.
\newblock {\em Nature}, 397(6714):41--43, 1999.

\bibitem{hurley2010new}
K~Hurley, Antonia Rowlinson, E~Bellm, D~Perley, Igor~G Mitrofanov, Dmitry~V Golovin, Alexander~S Kozyrev, Maxim~L Litvak, AB~Sanin, W~Boynton, et~al.
\newblock A new analysis of the short-duration, hard-spectrum grb 051103, a possible extragalactic soft gamma repeater giant flare.
\newblock {\em Monthly Notices of the Royal Astronomical Society}, 403(1):342--352, 2010.

\bibitem{2023Natur.619..487H}
N.~{Hurley-Walker}, N.~{Rea}, S.~J. {McSweeney}, B.~W. {Meyers}, E.~{Lenc}, I.~{Heywood}, S.~D. {Hyman}, Y.~P. {Men}, T.~E. {Clarke}, F.~{Coti Zelati}, D.~C. {Price}, C.~{Horv{\'a}th}, T.~J. {Galvin}, G.~E. {Anderson}, A.~{Bahramian}, E.~D. {Barr}, N.~D.~R. {Bhat}, M.~{Caleb}, M.~{Dall'Ora}, D.~{de Martino}, S.~{Giacintucci}, J.~S. {Morgan}, K.~M. {Rajwade}, B.~{Stappers}, and A.~{Williams}.
\newblock {A long-period radio transient active for three decades}.
\newblock {\em \nat}, 619(7970):487--490, July 2023.

\bibitem{2022Natur.601..526H}
N.~{Hurley-Walker}, X.~{Zhang}, A.~{Bahramian}, S.~J. {McSweeney}, T.~N. {O'Doherty}, P.~J. {Hancock}, J.~S. {Morgan}, G.~E. {Anderson}, G.~H. {Heald}, and T.~J. {Galvin}.
\newblock {A radio transient with unusually slow periodic emission}.
\newblock {\em \nat}, 601(7894):526--530, January 2022.

\bibitem{2006APh....26..155I}
{IceCube Collaboration}, A.~{Achterberg}, M.~{Ackermann}, J.~{Adams}, et~al.
\newblock {First year performance of the IceCube neutrino telescope}.
\newblock {\em Astroparticle Physics}, 26(3):155--173, October 2006.

\bibitem{2021NatAs...5..145I}
Andrei~P. {Igoshev}, Rainer {Hollerbach}, Toby {Wood}, and Konstantinos~N. {Gourgouliatos}.
\newblock {Strong toroidal magnetic fields required by quiescent X-ray emission of magnetars}.
\newblock {\em Nature Astronomy}, 5:145--149, January 2021.

\bibitem{2005ApJ...633.1013I}
Kunihito {Ioka}, Soebur {Razzaque}, Shiho {Kobayashi}, and Peter {M{\'e}sz{\'a}ros}.
\newblock {TeV-PeV Neutrinos from Giant Flares of Magnetars and the Case of SGR 1806-20}.
\newblock {\em \apj}, 633(2):1013--1017, November 2005.

\bibitem{2005ApJ...628L..53I}
G.~L. {Israel}, T.~{Belloni}, L.~{Stella}, Y.~{Rephaeli}, D.~E. {Gruber}, P.~{Casella}, S.~{Dall'Osso}, N.~{Rea}, M.~{Persic}, and R.~E. {Rothschild}.
\newblock {The Discovery of Rapid X-Ray Oscillations in the Tail of the SGR 1806-20 Hyperflare}.
\newblock {\em \apjl}, 628(1):L53--L56, July 2005.

\bibitem{2008ApJ...685.1114I}
G.~L. {Israel}, P.~{Romano}, V.~{Mangano}, S.~{Dall'Osso}, G.~{Chincarini}, L.~{Stella}, S.~{Campana}, T.~{Belloni}, G.~{Tagliaferri}, A.~J. {Blustin}, T.~{Sakamoto}, K.~{Hurley}, S.~{Zane}, A.~{Moretti}, D.~{Palmer}, C.~{Guidorzi}, D.~N. {Burrows}, N.~{Gehrels}, and H.~A. {Krimm}.
\newblock {A Swift Gaze into the 2006 March 29 Burst Forest of SGR 1900+14}.
\newblock {\em The Astrophysical Journal}, 685(2):1114--1128, October 2008.

\bibitem{kargaltsev2015SSRv}
Oleg {Kargaltsev}, Beno{\^\i}t {Cerutti}, Yuri {Lyubarsky}, and Edoardo {Striani}.
\newblock {Pulsar-Wind Nebulae. Recent Progress in Observations and Theory}.
\newblock {\em \ssr}, 191(1-4):391--439, October 2015.

\bibitem{kaspi03ApJ}
V.~M. {Kaspi}, F.~P. {Gavriil}, P.~M. {Woods}, J.~B. {Jensen}, M.~S.~E. {Roberts}, and D.~{Chakrabarty}.
\newblock {A Major Soft Gamma Repeater-like Outburst and Rotation Glitch in the No-longer-so-anomalous X-Ray Pulsar 1E 2259+586}.
\newblock {\em \apjl}, 588(2):L93--L96, May 2003.

\bibitem{kaspi2017}
Victoria~M. {Kaspi} and Andrei~M. {Beloborodov}.
\newblock {Magnetars}.
\newblock {\em \araa}, 55(1):261--301, August 2017.

\bibitem{2021NatAs...5..414K}
F.~{Kirsten}, M.~P. {Snelders}, M.~{Jenkins}, K.~{Nimmo}, J.~{van den Eijnden}, J.~W.~T. {Hessels}, M.~P. {Gawro{\'n}ski}, and J.~{Yang}.
\newblock {Detection of two bright radio bursts from magnetar SGR 1935 + 2154}.
\newblock {\em Nature Astronomy}, 5:414--422, April 2021.

\bibitem{kouvatsos2022detectability}
Nikolaos Kouvatsos, Paul~D Lasky, Ryan Quitzow-James, and Mairi Sakellariadou.
\newblock Detectability of the gravitational-wave background produced by magnetar giant flares.
\newblock {\em Physical Review D}, 106(6):063007, 2022.

\bibitem{kouveliotou87ApJ}
C.~{Kouveliotou}, J.~P. {Norris}, T.~L. {Cline}, B.~R. {Dennis}, U.~D. {Desai}, L.~E. {Orwig}, E.~E. {Fenimore}, R.~W. {Klebesadel}, J.~G. {Laros}, J.~L. {Atteia}, M.~{Boer}, K.~{Hurley}, M.~{Neil}, G.~{Vedrenne}, A.~V. {Kuznetsov}, R.~A. {Sunyaev}, and O.~V. {Terekhov}.
\newblock {SMM Hard X-Ray Observations of the Soft Gamma-Ray Repeater 1806-20}.
\newblock {\em \apjl}, 322:L21, November 1987.

\bibitem{kunjipurayil2022impact}
Athul Kunjipurayil, Tianqi Zhao, Bharat Kumar, Bijay~K Agrawal, and Madappa Prakash.
\newblock Impact of the equation of state on f-and p-mode oscillations of neutron stars.
\newblock {\em Physical Review D}, 106(6):063005, 2022.

\bibitem{2002ApJ...574L..29L}
R.~C. {Lamb}, D.~W. {Fox}, D.~J. {Macomb}, and T.~A. {Prince}.
\newblock {Discovery of a Possible Anomalous X-Ray Pulsar in the Small Magellanic Cloud}.
\newblock {\em \apjl}, 574(1):L29--L32, July 2002.

\bibitem{2015MNRAS.449.2047L}
S.~K. {Lander}, N.~{Andersson}, D.~{Antonopoulou}, and A.~L. {Watts}.
\newblock {Magnetically driven crustquakes in neutron stars}.
\newblock {\em \mnras}, 449(2):2047--2058, May 2015.

\bibitem{2020MNRAS.494.4838L}
S.~K. {Lander} and D.~I. {Jones}.
\newblock {Magnetar birth: rotation rates and gravitational-wave emission}.
\newblock {\em \mnras}, 494(4):4838--4847, June 2020.

\bibitem{laros87ApJ}
J.~G. {Laros}, E.~E. {Fenimore}, R.~W. {Klebesadel}, J.~L. {Atteia}, M.~{Boer}, K.~{Hurley}, M.~{Niel}, G.~{Vedrenne}, S.~R. {Kane}, C.~{Kouveliotou}, T.~L. {Cline}, B.~R. {Dennis}, U.~D. {Desai}, L.~E. {Orwig}, A.~V. {Kuznetsov}, R.~A. {Sunyaev}, and O.~V. {Terekhov}.
\newblock {A New Type of Repetitive Behavior in a High-Energy Transient}.
\newblock {\em \apjl}, 320:L111, September 1987.

\bibitem{levin10ApJ}
Lina {Levin}, Matthew {Bailes}, Samuel {Bates}, N.~D.~Ramesh {Bhat}, Marta {Burgay}, Sarah {Burke-Spolaor}, Nichi {D'Amico}, Simon {Johnston}, Michael {Keith}, Michael {Kramer}, Sabrina {Milia}, Andrea {Possenti}, Nanda {Rea}, Ben {Stappers}, and Willem {van Straten}.
\newblock {A Radio-loud Magnetar in X-ray Quiescence}.
\newblock {\em \apjl}, 721(1):L33--L37, September 2010.

\bibitem{liNatAs1935}
C.~K. {Li}, L.~{Lin}, S.~L. {Xiong}, M.~Y. {Ge}, X.~B. {Li}, T.~P. {Li}, F.~J. {Lu}, S.~N. {Zhang}, Y.~L. {Tuo}, Y.~{Nang}, B.~{Zhang}, S.~{Xiao}, Y.~{Chen}, L.~M. {Song}, Y.~P. {Xu}, C.~Z. {Liu}, S.~M. {Jia}, X.~L. {Cao}, J.~L. {Qu}, S.~{Zhang}, Y.~D. {Gu}, J.~Y. {Liao}, X.~F. {Zhao}, Y.~{Tan}, J.~Y. {Nie}, H.~S. {Zhao}, S.~J. {Zheng}, Y.~G. {Zheng}, Q.~{Luo}, C.~{Cai}, B.~{Li}, W.~C. {Xue}, Q.~C. {Bu}, Z.~{Chang}, G.~{Chen}, L.~{Chen}, T.~X. {Chen}, Y.~B. {Chen}, Y.~P. {Chen}, W.~{Cui}, W.~W. {Cui}, J.~K. {Deng}, Y.~W. {Dong}, Y.~Y. {Du}, M.~X. {Fu}, G.~H. {Gao}, H.~{Gao}, M.~{Gao}, Y.~D. {Gu}, J.~{Guan}, C.~C. {Guo}, D.~W. {Han}, Y.~{Huang}, J.~{Huo}, L.~H. {Jiang}, W.~C. {Jiang}, J.~{Jin}, Y.~J. {Jin}, L.~D. {Kong}, G.~{Li}, M.~S. {Li}, W.~{Li}, X.~{Li}, X.~F. {Li}, Y.~G. {Li}, Z.~W. {Li}, X.~H. {Liang}, B.~S. {Liu}, G.~Q. {Liu}, H.~W. {Liu}, X.~J. {Liu}, Y.~N. {Liu}, B.~{Lu}, X.~F. {Lu}, T.~{Luo}, X.~{Ma}, B.~{Meng}, G.~{Ou}, N.~{Sai}, R.~C. {Shang}, X.~Y. {Song}, L.~{Sun}, L.~{Tao}, C.~{Wang}, G.~F.
  {Wang}, J.~{Wang}, W.~S. {Wang}, Y.~S. {Wang}, X.~Y. {Wen}, B.~B. {Wu}, B.~Y. {Wu}, M.~{Wu}, G.~C. {Xiao}, H.~{Xu}, J.~W. {Yang}, S.~{Yang}, Y.~J. {Yang}, Yi-Jung {Yang}, Q.~B. {Yi}, Q.~Q. {Yin}, Y.~{You}, A.~M. {Zhang}, C.~M. {Zhang}, F.~{Zhang}, H.~M. {Zhang}, J.~{Zhang}, T.~{Zhang}, W.~{Zhang}, W.~C. {Zhang}, W.~Z. {Zhang}, Y.~{Zhang}, Yue {Zhang}, Y.~F. {Zhang}, Y.~J. {Zhang}, Z.~{Zhang}, Zhi {Zhang}, Z.~L. {Zhang}, D.~K. {Zhou}, J.~F. {Zhou}, Y.~{Zhu}, Y.~X. {Zhu}, and R.~L. {Zhuang}.
\newblock {HXMT identification of a non-thermal X-ray burst from SGR J1935+2154 and with FRB 200428}.
\newblock {\em Nature Astronomy}, 5:378--384, April 2021.

\bibitem{2022ApJ...931...56L}
Xiaobo {Li}, Mingyu {Ge}, Lin {Lin}, Shuang-Nan {Zhang}, Liming {Song}, Xuelei {Cao}, Bing {Zhang}, Fangjun {Lu}, Yupeng {Xu}, Shaolin {Xiong}, Youli {Tuo}, Ying {Tan}, Weichun {Jiang}, Jinlu {Qu}, Shu {Zhang}, Lingjun {Wang}, Jieshuang {Wang}, Binbin {Zhang}, Peng {Zhang}, Chengkui {Li}, Congzhan {Liu}, Tipei {Li}, Qingcui {Bu}, Ce~{Cai}, Yong {Chen}, Yupeng {Chen}, Zhi {Chang}, Li~{Chen}, Tianxiang {Chen}, Yibao {Chen}, Weiwei {Cui}, Yuanyuan {Du}, Guanhua {Gao}, He~{Gao}, Yudong {Gu}, Ju~{Guan}, Chengcheng {Guo}, Dawei {Han}, Yue {Huang}, Jia {Huo}, Shumei {Jia}, Jing {Jin}, Lingda {Kong}, Bing {Li}, Gang {Li}, Wei {Li}, Xian {Li}, Xufang {Li}, Zhengwei {Li}, Xiaohua {Liang}, Jinyuan {Liao}, Hexin {Liu}, Hongwei {Liu}, Xiaojing {Liu}, Xuefeng {Lu}, Qi~{Luo}, Tao {Luo}, Binyuan {Ma}, Ruican {Ma}, Xiang {Ma}, Bin {Meng}, Yi~{Nang}, Jianyin {Nie}, Ge~{Ou}, Xiaoqin {Ren}, Na~{Sai}, Xinying {Song}, Liang {Sun}, Lian {Tao}, Chen {Wang}, Pengju {Wang}, Wenshuai {Wang}, Yusa {Wang}, Xiangyang {Wen}, Bobing {Wu},
  Baiyang {Wu}, Mei {Wu}, Shuo {Xiao}, Sheng {Yang}, Yanji {Yang}, Qibin {Yi}, Qianqing {Yin}, Yuan {You}, Wei {Yu}, Fan {Zhang}, Hongmei {Zhang}, Juan {Zhang}, Wanchang {Zhang}, Wei {Zhang}, Yifei {Zhang}, Yuanhang {Zhang}, Haisheng {Zhao}, Xiaofan {Zhao}, Shijie {Zheng}, and Dengke {Zhou}.
\newblock {Quasi-periodic Oscillations of the X-Ray Burst from the Magnetar SGR J1935-2154 and Associated with the Fast Radio Burst FRB 200428}.
\newblock {\em \apj}, 931(1):56, May 2022.

\bibitem{2018NatCo...9.3833L}
Zheng-Xiang {Li}, He~{Gao}, Xu-Heng {Ding}, Guo-Jian {Wang}, and Bing {Zhang}.
\newblock {Strongly lensed repeating fast radio bursts as precision probes of the universe}.
\newblock {\em Nature Communications}, 9:3833, September 2018.

\bibitem{2017PhRvLnsm}
{LIGO Scientific Collaboration} and {Virgo Collaboration}.
\newblock {GW170817: Observation of Gravitational Waves from a Binary Neutron Star Inspiral}.
\newblock {\em \prl}, 119(16):161101, October 2017.

\bibitem{Linscott1980}
I.~R. {Linscott} and J.~W. {Erkes}.
\newblock {Discovery of millisecond radio bursts from M 87}.
\newblock {\em \apjl}, 236:L109--L113, March 1980.

\bibitem{lorimer2007bright}
Duncan~R Lorimer, Matthew Bailes, Maura~Ann McLaughlin, David~J Narkevic, and Froney Crawford.
\newblock A bright millisecond radio burst of extragalactic origin.
\newblock {\em Science}, 318(5851):777--780, 2007.

\bibitem{lower2020}
M.~E. {Lower}, M.~{Bailes}, R.~M. {Shannon}, S.~{Johnston}, C.~{Flynn}, S.~{Os{\l}owski}, V.~{Gupta}, W.~{Farah}, T.~{Bateman}, A.~J. {Green}, R.~{Hunstead}, A.~{Jameson}, F.~{Jankowski}, A.~{Parthasarathy}, D.~C. {Price}, A.~{Sutherland}, D.~{Temby}, and V.~{Venkatraman Krishnan}.
\newblock {The UTMOST pulsar timing programme - II. Timing noise across the pulsar population}.
\newblock {\em \mnras}, 494(1):228--245, May 2020.

\bibitem{lower20ApJ:1818}
Marcus~E. {Lower}, Ryan~M. {Shannon}, Simon {Johnston}, and Matthew {Bailes}.
\newblock {Spectropolarimetric Properties of Swift J1818.0-1607: A 1.4 s Radio Magnetar}.
\newblock {\em \apjl}, 896(2):L37, June 2020.

\bibitem{lower23ApJ}
Marcus~E. {Lower}, George {Younes}, Paul {Scholz}, Fernando {Camilo}, Liam {Dunn}, Simon {Johnston}, Teruaki {Enoto}, John~M. {Sarkissian}, John~E. {Reynolds}, David~M. {Palmer}, Zaven {Arzoumanian}, Matthew~G. {Baring}, Keith {Gendreau}, Ersin {G{\"o}{\u{g}}{\"u}{\c{s}}}, Sebastien {Guillot}, Alexander~J. {van der Horst}, Chin-Ping {Hu}, Chryssa {Kouveliotou}, Lin {Lin}, Christian {Malacaria}, Rachael {Stewart}, and Zorawar {Wadiasingh}.
\newblock {The 2022 High-energy Outburst and Radio Disappearing Act of the Magnetar 1E 1547.0-5408}.
\newblock {\em \apj}, 945(2):153, March 2023.

\bibitem{2022ATel15697....1M}
Yogesh {Maan}, Joeri~van {Leeuwen}, Samayra {Straal}, and Ines {Pastor-Marazuela}.
\newblock {GBT detection of bright 5 GHz radio bursts from SGR 1935+2154, coincident with X-ray and 600 MHz bursts}.
\newblock {\em The Astronomer's Telegram}, 15697:1, October 2022.

\bibitem{Macquart20}
J.~P. {Macquart}, J.~X. {Prochaska}, M.~{McQuinn}, K.~W. {Bannister}, S.~{Bhandari}, C.~K. {Day}, A.~T. {Deller}, R.~D. {Ekers}, C.~W. {James}, L.~{Marnoch}, S.~{Os{\l}owski}, C.~{Phillips}, S.~D. {Ryder}, D.~R. {Scott}, R.~M. {Shannon}, and N.~{Tejos}.
\newblock {A census of baryons in the Universe from localized fast radio bursts}.
\newblock {\em \nat}, 581(7809):391--395, May 2020.

\bibitem{Macquet:2021eyn}
Adrian Macquet, Marie-Anne Bizouard, Eric Burns, Nelson Christensen, Michael Coughlin, Zorawar Wadiasingh, and George Younes.
\newblock {Search for Long-duration Gravitational-wave Signals Associated with Magnetar Giant Flares}.
\newblock {\em Astrophys. J.}, 918(2):80, 2021.

\bibitem{2005AJ....129.1993M}
R.~N. {Manchester}, G.~B. {Hobbs}, A.~{Teoh}, and M.~{Hobbs}.
\newblock {The Australia Telescope National Facility Pulsar Catalogue}.
\newblock {\em \aj}, 129(4):1993--2006, April 2005.

\bibitem{2023HEAD...2010367M}
Herman {Marshall}, Sarah {Heine}, Alan {Garner}, Stephen {Bongiorno}, Hans~Moritz {Guenther}, Ralf {Heilmann}, Rebecca {Masterson}, and Alan {Marscher}.
\newblock {The Rocket Experiment Demonstration of a Soft X-ray Polarimeter}.
\newblock In {\em AAS/High Energy Astrophysics Division}, volume~55 of {\em AAS/High Energy Astrophysics Division}, page 103.67, September 2023.

\bibitem{mazets2008giant}
EP~Mazets, RL~Aptekar, TL~Cline, DD~Frederiks, JO~Goldsten, SV~Golenetskii, K~Hurley, A~Von~Kienlin, and VD~Palshin.
\newblock A giant flare from a soft gamma repeater in the andromeda galaxy (m31).
\newblock {\em The Astrophysical Journal}, 680(1):545, 2008.

\bibitem{mazets1979observations}
EP~Mazets, SV~Golenetskii, VN~Il'Inskii, RL~Aptekar', and Yu~A Guryan.
\newblock Observations of a flaring x-ray pulsar in dorado.
\newblock {\em Nature}, 282:587--589, 1979.

\bibitem{meegan2009fermi}
Charles Meegan, Giselher Lichti, PN~Bhat, Elisabetta Bissaldi, Michael~S Briggs, Valerie Connaughton, Roland Diehl, Gerald Fishman, Jochen Greiner, Andrew~S Hoover, et~al.
\newblock The fermi gamma-ray burst monitor.
\newblock {\em The Astrophysical Journal}, 702(1):791, 2009.

\bibitem{mereghetti20ApJ}
S.~{Mereghetti}, V.~{Savchenko}, C.~{Ferrigno}, D.~{G{\"o}tz}, M.~{Rigoselli}, A.~{Tiengo}, A.~{Bazzano}, E.~{Bozzo}, A.~{Coleiro}, T.~J.~L. {Courvoisier}, M.~{Doyle}, A.~{Goldwurm}, L.~{Hanlon}, E.~{Jourdain}, A.~{von Kienlin}, A.~{Lutovinov}, A.~{Martin-Carrillo}, S.~{Molkov}, L.~{Natalucci}, F.~{Onori}, F.~{Panessa}, J.~{Rodi}, J.~{Rodriguez}, C.~{S{\'a}nchez-Fern{\'a}ndez}, R.~{Sunyaev}, and P.~{Ubertini}.
\newblock {INTEGRAL Discovery of a Burst with Associated Radio Emission from the Magnetar SGR 1935+2154}.
\newblock {\em \apjl}, 898(2):L29, August 2020.

\bibitem{2008AandARv..15..225M}
Sandro {Mereghetti}.
\newblock {The strongest cosmic magnets: soft gamma-ray repeaters and anomalous X-ray pulsars}.
\newblock {\em \aapr}, 15(4):225--287, July 2008.

\bibitem{2015SSRv..191..315M}
Sandro {Mereghetti}, Jos{\'e}~A. {Pons}, and Andrew {Melatos}.
\newblock {Magnetars: Properties, Origin and Evolution}.
\newblock {\em \ssr}, 191(1-4):315--338, October 2015.

\bibitem{mereghetti2023magnetar}
Sandro Mereghetti, Michela Rigoselli, Ruben Salvaterra, Dominik~P Pacholski, James~C Rodi, Diego Gotz, Edoardo Arrigoni, Paolo d'Avanzo, Christophe Adami, Angela Bazzano, et~al.
\newblock A magnetar giant flare in the nearby starburst galaxy m82.
\newblock {\em arXiv preprint arXiv:2312.14645}, 2023.

\bibitem{2019MNRAS.485.4091M}
Brian~D. {Metzger}, Ben {Margalit}, and Lorenzo {Sironi}.
\newblock {Fast radio bursts as synchrotron maser emission from decelerating relativistic blast waves}.
\newblock {\em \mnras}, 485(3):4091--4106, May 2019.

\bibitem{miller19ApJ}
M.~C. {Miller}, F.~K. {Lamb}, A.~J. {Dittmann}, S.~{Bogdanov}, Z.~{Arzoumanian}, K.~C. {Gendreau}, S.~{Guillot}, A.~K. {Harding}, W.~C.~G. {Ho}, J.~M. {Lattimer}, R.~M. {Ludlam}, S.~{Mahmoodifar}, S.~M. {Morsink}, P.~S. {Ray}, T.~E. {Strohmayer}, K.~S. {Wood}, T.~{Enoto}, R.~{Foster}, T.~{Okajima}, G.~{Prigozhin}, and Y.~{Soong}.
\newblock {PSR J0030+0451 Mass and Radius from NICER Data and Implications for the Properties of Neutron Star Matter}.
\newblock {\em \apjl}, 887(1):L24, December 2019.

\bibitem{2023IAUS..363..284N}
Michela {Negro} and Eric {Burns}.
\newblock {Identification of a Local Sample of Gamma-Ray Bursts Consistent with a Magnetar Giant Flare Origin}.
\newblock {\em IAU Symposium}, 363:284--287, January 2023.

\bibitem{negro2023ixpe}
Michela Negro, Niccol{\`o} Di~Lalla, Nicola Omodei, P{\'e}ter Veres, Stefano Silvestri, Alberto Manfreda, Eric Burns, Luca Baldini, Enrico Costa, Steven~R Ehlert, et~al.
\newblock The ixpe view of grb 221009a.
\newblock {\em The Astrophysical Journal Letters}, 946(1):L21, 2023.

\bibitem{Note1}
eROSITA might detect few magnetars at the end of its full-sky survey, yet these will likely be marked as candidates as many might not be bright enough for pulsation detection.

\bibitem{Note2}
Although possibly much earlier by \cite {Linscott1980}.

\bibitem{2008ApJ...681.1464O}
E.~O. {Ofek}, M.~{Muno}, R.~{Quimby}, S.~R. {Kulkarni}, H.~{Stiele}, W.~{Pietsch}, E.~{Nakar}, A.~{Gal-Yam}, A.~{Rau}, P.~B. {Cameron}, S.~B. {Cenko}, M.~M. {Kasliwal}, D.~B. {Fox}, P.~{Chandra}, A.~K.~H. {Kong}, and R.~{Barnard}.
\newblock {GRB 070201: A Possible Soft Gamma-Ray Repeater in M31}.
\newblock {\em \apj}, 681(2):1464--1469, July 2008.

\bibitem{ofek2007soft}
Eran~O Ofek.
\newblock Soft gamma-ray repeaters in nearby galaxies: rate, luminosity function, and fraction among short gamma-ray bursts.
\newblock {\em The Astrophysical Journal}, 659(1):339, 2007.

\bibitem{2014ApJS..212....6O}
S.~A. {Olausen} and V.~M. {Kaspi}.
\newblock {The McGill Magnetar Catalog}.
\newblock {\em \apjs}, 212(1):6, May 2014.

\bibitem{Paczynski1992}
Bohdan {Paczynski}.
\newblock {GB 790305 as a Very Strongly Magnetized Neutron Star}.
\newblock {\em \actaa}, 42:145--153, July 1992.

\bibitem{2020ATel13675....1P}
David~M. {Palmer}.
\newblock {A Forest of Bursts from SGR 1935+2154}.
\newblock {\em The Astronomer's Telegram}, 13675:1, April 2020.

\bibitem{palmer2005giant}
David~M Palmer, S~Barthelmy, N~Gehrels, RM~Kippen, T~Cayton, C~Kouveliotou, D~Eichler, RAMJ Wijers, PM~Woods, J~Granot, et~al.
\newblock A giant $\gamma$-ray flare from the magnetar sgr 1806--20.
\newblock {\em Nature}, 434(7037):1107--1109, 2005.

\bibitem{pearlman23arXiv230810930P}
Aaron~B. {Pearlman}, Paul {Scholz}, Suryarao {Bethapudi}, Jason W.~T. {Hessels}, Victoria~M. {Kaspi}, Franz {Kirsten}, Kenzie {Nimmo}, Laura~G. {Spitler}, Emmanuel {Fonseca}, Bradley~W. {Meyers}, Ingrid {Stairs}, Chia~Min {Tan}, Mohit {Bhardwaj}, Shami {Chatterjee}, Amanda~M. {Cook}, Alice~P. {Curtin}, Fengqiu~Adam {Dong}, Tarraneh {Eftekhari}, B.~M. {Gaensler}, Tolga {G{\"u}ver}, Jane {Kaczmarek}, Calvin {Leung}, Kiyoshi~W. {Masui}, Daniele {Michilli}, Thomas~A. {Prince}, Ketan~R. {Sand}, Kaitlyn {Shin}, Kendrick~M. {Smith}, and Shriharsh~P. {Tendulkar}.
\newblock {Multiwavelength Constraints on the Origin of a Nearby Repeating Fast Radio Burst Source in a Globular Cluster}.
\newblock {\em arXiv e-prints}, page arXiv:2308.10930, August 2023.

\bibitem{perna11ApJ}
Rosalba {Perna} and Jose~A. {Pons}.
\newblock {A Unified Model of the Magnetar and Radio Pulsar Bursting Phenomenology}.
\newblock {\em \apjl}, 727(2):L51, February 2011.

\bibitem{Platts+19}
E.~{Platts}, A.~{Weltman}, A.~{Walters}, S.~P. {Tendulkar}, J.~E.~B. {Gordin}, and S.~{Kandhai}.
\newblock {A living theory catalogue for fast radio bursts}.
\newblock {\em \physrep}, 821:1--27, August 2019.

\bibitem{pons09AA}
J.~A. {Pons}, J.~A. {Miralles}, and U.~{Geppert}.
\newblock {Magneto-thermal evolution of neutron stars}.
\newblock {\em \aap}, 496(1):207--216, March 2009.

\bibitem{Popov&Postnov13}
S.~B. {Popov} and K.~A. {Postnov}.
\newblock {Millisecond extragalactic radio bursts as magnetar flares}.
\newblock {\em arXiv e-prints}, page arXiv:1307.4924, July 2013.

\bibitem{Popov_2018}
S~B Popov, K~A Postnov, and M~S Pshirkov.
\newblock Fast radio bursts.
\newblock {\em Physics-Uspekhi}, 61(10):965, oct 2018.

\bibitem{2010vaoa.conf..129P}
Sergey~B. {Popov} and K.~A. {Postnov}.
\newblock {Hyperflares of SGRs as an engine for millisecond extragalactic radio bursts}.
\newblock In H.~A. {Harutyunian}, A.~M. {Mickaelian}, and Y.~{Terzian}, editors, {\em Evolution of Cosmic Objects through their Physical Activity}, pages 129--132, November 2010.

\bibitem{2021A&A...647A...1P}
P.~{Predehl}, R.~{Andritschke}, V.~{Arefiev}, et~al.
\newblock {The eROSITA X-ray telescope on SRG}.
\newblock {\em \aap}, 647:A1, March 2021.

\bibitem{2020MNRAS.495.3551R}
K.~M. {Rajwade}, M.~B. {Mickaliger}, B.~W. {Stappers}, V.~{Morello}, D.~{Agarwal}, C.~G. {Bassa}, R.~P. {Breton}, M.~{Caleb}, A.~{Karastergiou}, E.~F. {Keane}, and D.~R. {Lorimer}.
\newblock {Possible periodic activity in the repeating FRB 121102}.
\newblock {\em \mnras}, 495(4):3551--3558, July 2020.

\bibitem{2020ATel14112....1R}
P.~S. {Ray}, G.~{Younes}, T.~{Guver}, Zorawar {Wadiasingh}, Z.~{Arzoumanian}, K.~C. {Gendreau}, T.~{Enoto}, W.~C.~G. {Ho}, C.~P. {Hu}, K.~{Bansal}, and W.~{Majid}.
\newblock {NICER confirmation of the magnetar nature of SGR 1830-0645: Spindown, spectra, and short bursts}.
\newblock {\em The Astronomer's Telegram}, 14112:1, October 2020.

\bibitem{rea2016ApJ}
N.~{Rea}, A.~{Borghese}, P.~{Esposito}, F.~{Coti Zelati}, M.~{Bachetti}, G.~L. {Israel}, and A.~{De Luca}.
\newblock {Magnetar-like Activity from the Central Compact Object in the SNR RCW103}.
\newblock {\em \apjl}, 828(1):L13, September 2016.

\bibitem{2022ApJ...940...72R}
N.~{Rea}, F.~{Coti Zelati}, C.~{Dehman}, N.~{Hurley-Walker}, D.~{de Martino}, A.~{Bahramian}, D.~A.~H. {Buckley}, J.~{Brink}, A.~{Kawka}, J.~A. {Pons}, D.~{Vigan{\`o}}, V.~{Graber}, M.~{Ronchi}, C.~{Pardo Araujo}, A.~{Borghese}, E.~{Parent}, and T.~J. {Galvin}.
\newblock {Constraining the Nature of the 18 min Periodic Radio Transient GLEAM-X J162759.5-523504.3 via Multiwavelength Observations and Magneto-thermal Simulations}.
\newblock {\em \apj}, 940(1):72, November 2022.

\bibitem{rea2010Sci}
N.~{Rea}, P.~{Esposito}, R.~{Turolla}, G.~L. {Israel}, S.~{Zane}, L.~{Stella}, S.~{Mereghetti}, A.~{Tiengo}, D.~{G{\"o}tz}, E.~{G{\"o}{\u{g}}{\"u}{\c{s}}}, and C.~{Kouveliotou}.
\newblock {A Low-Magnetic-Field Soft Gamma Repeater}.
\newblock {\em Science}, 330(6006):944, November 2010.

\bibitem{2024ApJ...961..214R}
N.~{Rea}, N.~{Hurley-Walker}, C.~{Pardo-Araujo}, M.~{Ronchi}, V.~{Graber}, F.~{Coti Zelati}, D.~{de Martino}, A.~{Bahramian}, S.~J. {McSweeney}, T.~J. {Galvin}, S.~D. {Hyman}, and M.~{Dall'Ora}.
\newblock {Long-period Radio Pulsars: Population Study in the Neutron Star and White Dwarf Rotating Dipole Scenarios}.
\newblock {\em \apj}, 961(2):214, February 2024.

\bibitem{rea09MNRAS:0501}
N.~{Rea}, G.~L. {Israel}, R.~{Turolla}, P.~{Esposito}, S.~{Mereghetti}, D.~{G{\"o}tz}, S.~{Zane}, A.~{Tiengo}, K.~{Hurley}, M.~{Feroci}, M.~{Still}, V.~{Yershov}, C.~{Winkler}, R.~{Perna}, F.~{Bernardini}, P.~{Ubertini}, L.~{Stella}, S.~{Campana}, M.~{van der Klis}, and P.~{Woods}.
\newblock {The first outburst of the new magnetar candidate SGR0501+4516}.
\newblock {\em \mnras}, 396(4):2419--2432, July 2009.

\bibitem{ridnaia21NatAs}
A.~{Ridnaia}, D.~{Svinkin}, D.~{Frederiks}, A.~{Bykov}, S.~{Popov}, R.~{Aptekar}, S.~{Golenetskii}, A.~{Lysenko}, A.~{Tsvetkova}, M.~{Ulanov}, and T.~L. {Cline}.
\newblock {A peculiar hard X-ray counterpart of a Galactic fast radio burst}.
\newblock {\em Nature Astronomy}, 5:372--377, April 2021.

\bibitem{riley19ApJ}
T.~E. {Riley}, A.~L. {Watts}, S.~{Bogdanov}, P.~S. {Ray}, R.~M. {Ludlam}, S.~{Guillot}, Z.~{Arzoumanian}, C.~L. {Baker}, A.~V. {Bilous}, D.~{Chakrabarty}, K.~C. {Gendreau}, A.~K. {Harding}, W.~C.~G. {Ho}, J.~M. {Lattimer}, S.~M. {Morsink}, and T.~E. {Strohmayer}.
\newblock {A NICER View of PSR J0030+0451: Millisecond Pulsar Parameter Estimation}.
\newblock {\em \apjl}, 887(1):L21, December 2019.

\bibitem{2021Natur.589..207R}
O.~J. {Roberts}, P.~{Veres}, M.~G. {Baring}, M.~S. {Briggs}, C.~{Kouveliotou}, E.~{Bissaldi}, G.~{Younes}, S.~I. {Chastain}, J.~J. {DeLaunay}, D.~{Huppenkothen}, A.~{Tohuvavohu}, P.~N. {Bhat}, E.~{G{\"o}{\v{g}}{\"u}{\c{s}}}, A.~J. {van der Horst}, J.~A. {Kennea}, D.~{Kocevski}, J.~D. {Linford}, S.~{Guiriec}, R.~{Hamburg}, C.~A. {Wilson-Hodge}, and E.~{Burns}.
\newblock {Rapid spectral variability of a giant flare from a magnetar in NGC 253}.
\newblock {\em \nat}, 589(7841):207--210, January 2021.

\bibitem{2012A&A...541A.122S}
V.~{Savchenko}, A.~{Neronov}, and T.~J.~L. {Courvoisier}.
\newblock {Timing properties of gamma-ray bursts detected by SPI-ACS detector onboard INTEGRAL}.
\newblock {\em \aap}, 541:A122, May 2012.

\bibitem{2019Natur.574..211S}
Fabian R.~N. {Schneider}, Sebastian~T. {Ohlmann}, Philipp {Podsiadlowski}, Friedrich~K. {R{\"o}pke}, Steven~A. {Balbus}, R{\"u}diger {Pakmor}, and Volker {Springel}.
\newblock {Stellar mergers as the origin of magnetic massive stars}.
\newblock {\em \nat}, 574(7777):211--214, October 2019.

\bibitem{2014ApJ...790..101S}
L.~G. {Spitler}, J.~M. {Cordes}, J.~W.~T. {Hessels}, D.~R. {Lorimer}, M.~A. {McLaughlin}, S.~{Chatterjee}, F.~{Crawford}, J.~S. {Deneva}, V.~M. {Kaspi}, R.~S. {Wharton}, B.~{Allen}, S.~{Bogdanov}, A.~{Brazier}, F.~{Camilo}, P.~C.~C. {Freire}, F.~A. {Jenet}, C.~{Karako-Argaman}, B.~{Knispel}, P.~{Lazarus}, K.~J. {Lee}, J.~{van Leeuwen}, R.~{Lynch}, S.~M. {Ransom}, P.~{Scholz}, X.~{Siemens}, I.~H. {Stairs}, K.~{Stovall}, J.~K. {Swiggum}, A.~{Venkataraman}, W.~W. {Zhu}, C.~{Aulbert}, and H.~{Fehrmann}.
\newblock {Fast Radio Burst Discovered in the Arecibo Pulsar ALFA Survey}.
\newblock {\em \apj}, 790(2):101, August 2014.

\bibitem{2016Natur.531..202S}
L.~G. {Spitler}, P.~{Scholz}, J.~W.~T. {Hessels}, S.~{Bogdanov}, A.~{Brazier}, F.~{Camilo}, S.~{Chatterjee}, J.~M. {Cordes}, F.~{Crawford}, J.~{Deneva}, R.~D. {Ferdman}, P.~C.~C. {Freire}, V.~M. {Kaspi}, P.~{Lazarus}, R.~{Lynch}, E.~C. {Madsen}, M.~A. {McLaughlin}, C.~{Patel}, S.~M. {Ransom}, A.~{Seymour}, I.~H. {Stairs}, B.~W. {Stappers}, J.~{van Leeuwen}, and W.~W. {Zhu}.
\newblock {A repeating fast radio burst}.
\newblock {\em \nat}, 531(7593):202--205, March 2016.

\bibitem{2005ApJ...634L.165S}
L.~{Stella}, S.~{Dall'Osso}, G.~L. {Israel}, and A.~{Vecchio}.
\newblock {Gravitational Radiation from Newborn Magnetars in the Virgo Cluster}.
\newblock {\em \apjl}, 634(2):L165--L168, December 2005.

\bibitem{2005ApJ...632L.111S}
Tod~E. {Strohmayer} and Anna~L. {Watts}.
\newblock {Discovery of Fast X-Ray Oscillations during the 1998 Giant Flare from SGR 1900+14}.
\newblock {\em \apjl}, 632(2):L111--L114, October 2005.

\bibitem{2006ApJ...653..593S}
Tod~E. {Strohmayer} and Anna~L. {Watts}.
\newblock {The 2004 Hyperflare from SGR 1806-20: Further Evidence for Global Torsional Vibrations}.
\newblock {\em \apj}, 653(1):593--601, December 2006.

\bibitem{svinkin2021bright}
D~Svinkin, D~Frederiks, K~Hurley, R~Aptekar, S~Golenetskii, A~Lysenko, AV~Ridnaia, A~Tsvetkova, M~Ulanov, TL~Cline, et~al.
\newblock A bright $\gamma$-ray flare interpreted as a giant magnetar flare in ngc 253.
\newblock {\em Nature}, 589(7841):211--213, 2021.

\bibitem{1999NuPhS..69...12S}
J.~H. {Swank}.
\newblock {The Rossi X-Ray Timing Explorer}.
\newblock {\em Nuclear Physics B Proceedings Supplements}, 69(1-3):12--19, January 1999.

\bibitem{tam04ApJ:2259}
Cindy~R. {Tam}, Victoria~M. {Kaspi}, Marten~H. {van Kerkwijk}, and Martin {Durant}.
\newblock {Correlated Infrared and X-Ray Flux Changes Following the 2002 June Outburst of the Anomalous X-Ray Pulsar 1E 2259+586}.
\newblock {\em \apjl}, 617(1):L53--L56, December 2004.

\bibitem{tavani21NatAs}
M.~{Tavani}, C.~{Casentini}, A.~{Ursi}, F.~{Verrecchia}, A.~{Addis}, L.~A. {Antonelli}, A.~{Argan}, G.~{Barbiellini}, L.~{Baroncelli}, G.~{Bernardi}, G.~{Bianchi}, A.~{Bulgarelli}, P.~{Caraveo}, M.~{Cardillo}, P.~W. {Cattaneo}, A.~W. {Chen}, E.~{Costa}, E.~{Del Monte}, G.~{Di Cocco}, G.~{Di Persio}, I.~{Donnarumma}, Y.~{Evangelista}, M.~{Feroci}, A.~{Ferrari}, V.~{Fioretti}, F.~{Fuschino}, M.~{Galli}, F.~{Gianotti}, A.~{Giuliani}, C.~{Labanti}, F.~{Lazzarotto}, P.~{Lipari}, F.~{Longo}, F.~{Lucarelli}, A.~{Magro}, M.~{Marisaldi}, S.~{Mereghetti}, E.~{Morelli}, A.~{Morselli}, G.~{Naldi}, L.~{Pacciani}, N.~{Parmiggiani}, F.~{Paoletti}, A.~{Pellizzoni}, M.~{Perri}, F.~{Perotti}, G.~{Piano}, P.~{Picozza}, M.~{Pilia}, C.~{Pittori}, S.~{Puccetti}, G.~{Pupillo}, M.~{Rapisarda}, A.~{Rappoldi}, A.~{Rubini}, G.~{Setti}, P.~{Soffitta}, M.~{Trifoglio}, A.~{Trois}, S.~{Vercellone}, V.~{Vittorini}, P.~{Giommi}, and F.~{D'Amico}.
\newblock {An X-ray burst from a magnetar enlightening the mechanism of fast radio bursts}.
\newblock {\em Nature Astronomy}, 5:401--407, April 2021.

\bibitem{2017MNRAS.469.3610T}
R.~{Taverna} and R.~{Turolla}.
\newblock {On the spectrum and polarization of magnetar flare emission}.
\newblock {\em \mnras}, 469(3):3610--3628, August 2017.

\bibitem{2016ApJ...827...59T}
Shriharsh~P. {Tendulkar}, Victoria~M. {Kaspi}, and Chitrang {Patel}.
\newblock {Radio Nondetection of the SGR 1806-20 Giant Flare and Implications for Fast Radio Bursts}.
\newblock {\em \apj}, 827(1):59, August 2016.

\bibitem{2023arXiv230812362T}
{The COSI COllaboration}, John~A. {Tomsick}, Steven~E. {Boggs}, Andreas {Zoglauer}, et~al.
\newblock {The Compton Spectrometer and Imager}.
\newblock {\em arXiv e-prints}, page arXiv:2308.12362, August 2023.

\bibitem{LATMGF}
{The Fermi-LAT Collaboration}, M.~{Ajello}, W.~B. {Atwood}, M.~{Axelsson}, et~al.
\newblock {High-energy emission from a magnetar giant flare in the Sculptor galaxy}.
\newblock {\em Nature Astronomy}, 5:385--391, April 2021.

\bibitem{IceCube}
{The IceCube Collaboration}, M.~G. {Aartsen}, M.~{Ackermann}, J.~{Adams}, et~al.
\newblock {The IceCube Neutrino Observatory: instrumentation and online systems}.
\newblock {\em Journal of Instrumentation}, 12(3):P03012, March 2017.

\bibitem{2022Sci...378..646T}
{The IXPE Collaboration}, Roberto {Taverna}, Roberto {Turolla}, Fabio {Muleri}, et~al.
\newblock {Polarized x-rays from a magnetar}.
\newblock {\em Science}, 378(6620):646--650, November 2022.

\bibitem{2023ApJ...944L..27Z}
{The IXPE Collaboration}, Silvia {Zane}, Roberto {Taverna}, Denis {Gonz{\'a}lez-Caniulef}, et~al.
\newblock {A Strong X-Ray Polarization Signal from the Magnetar 1RXS J170849.0-400910}.
\newblock {\em \apjl}, 944(2):L27, February 2023.

\bibitem{2022JATIS...8b6002W}
Martin~C. {The IXPE Collaboration}, and{Weisskopf}, Paolo {Soffitta}, Luca {Baldini}, et~al.
\newblock {The Imaging X-Ray Polarimetry Explorer (IXPE): Pre-Launch}.
\newblock {\em Journal of Astronomical Telescopes, Instruments, and Systems}, 8(2):026002, April 2022.

\bibitem{O3LVKmagnetar}
{The LIGO Scientific Collaboration}, {the Virgo Collaboration}, and {the KAGRA Collaboration}.
\newblock {Search for gravitational-wave transients associated with magnetar bursts in Advanced LIGO and Advanced Virgo data from the third observing run}.
\newblock {\em arXiv e-prints}, page arXiv:2210.10931, October 2022.

\bibitem{TD1993}
Christopher {Thompson} and Robert~C. {Duncan}.
\newblock {Neutron Star Dynamos and the Origins of Pulsar Magnetism}.
\newblock {\em \apj}, 408:194, May 1993.

\bibitem{thompson96ApJ}
Christopher {Thompson} and Robert~C. {Duncan}.
\newblock {The Soft Gamma Repeaters as Very Strongly Magnetized Neutron Stars. II. Quiescent Neutrino, X-Ray, and Alfven Wave Emission}.
\newblock {\em \apj}, 473:322, December 1996.

\bibitem{2023MNRAS.526.2795T}
Tomonori {Totani} and Yuya {Tsuzuki}.
\newblock {Fast radio bursts trigger aftershocks resembling earthquakes, but not solar flares}.
\newblock {\em \mnras}, 526(2):2795--2811, December 2023.

\bibitem{trigg2023grb}
Aaron~C Trigg, Eric Burns, Oliver~J Roberts, Michela Negro, Dmitry~S Svinkin, Matthew~G Baring, Zorawar Wadiasingh, Nelson~L Christensen, Igor Andreoni, Michael~S Briggs, et~al.
\newblock Grb 180128a: A second magnetar giant flare candidate from the sculptor galaxy.
\newblock {\em arXiv preprint arXiv:2311.09362}, 2023.

\bibitem{1982AdSpR...2d.241T}
J.~{Truemper}.
\newblock {The ROSAT mission}.
\newblock {\em Advances in Space Research}, 2(4):241--249, January 1982.

\bibitem{2024arXiv240116758T}
Yuya {Tsuzuki}, Tomonori {Totani}, Chin-Ping {Hu}, and Teruaki {Enoto}.
\newblock {Similarity to earthquakes again: periodic radio pulses of the magnetar SGR 1935+2154 are accompanied by aftershocks like fast radio bursts}.
\newblock {\em arXiv e-prints}, page arXiv:2401.16758, January 2024.

\bibitem{2015RPPh...78k6901T}
R.~{Turolla}, S.~{Zane}, and A.~L. {Watts}.
\newblock {Magnetars: the physics behind observations. A review}.
\newblock {\em Reports on Progress in Physics}, 78(11):116901, November 2015.

\bibitem{2012ApJ...749..122V}
A.~J. {van der Horst}, C.~{Kouveliotou}, N.~M. {Gorgone}, Y.~{Kaneko}, M.~G. {Baring}, S.~{Guiriec}, E.~{G{\"o}{\v{g}}{\"u}{\c{s}}}, J.~{Granot}, A.~L. {Watts}, L.~{Lin}, P.~N. {Bhat}, E.~{Bissaldi}, V.~L. {Chaplin}, M.~H. {Finger}, N.~{Gehrels}, M.~H. {Gibby}, M.~M. {Giles}, A.~{Goldstein}, D.~{Gruber}, A.~K. {Harding}, L.~{Kaper}, A.~{von Kienlin}, M.~{van der Klis}, S.~{McBreen}, J.~{Mcenery}, C.~A. {Meegan}, W.~S. {Paciesas}, A.~{Pe'er}, R.~D. {Preece}, E.~{Ramirez-Ruiz}, A.~{Rau}, S.~{Wachter}, C.~{Wilson-Hodge}, P.~M. {Woods}, and R.~A.~M.~J. {Wijers}.
\newblock {SGR J1550-5418 Bursts Detected with the Fermi Gamma-Ray Burst Monitor during its Most Prolific Activity}.
\newblock {\em \apj}, 749(2):122, April 2012.

\bibitem{2016MNRAS.461..877V}
T.~{van Putten}, A.~L. {Watts}, M.~G. {Baring}, and R.~A.~M.~J. {Wijers}.
\newblock {Radiative transfer simulations of magnetar flare beaming}.
\newblock {\em \mnras}, 461(1):877--891, September 2016.

\bibitem{Vigano2013}
D.~{Vigan{\`o}}, N.~{Rea}, J.~A. {Pons}, R.~{Perna}, D.~N. {Aguilera}, and J.~A. {Miralles}.
\newblock {Unifying the observational diversity of isolated neutron stars via magneto-thermal evolution models}.
\newblock {\em \mnras}, 434(1):123--141, September 2013.

\bibitem{2020ApJ...891...82W}
Zorawar {Wadiasingh}, Paz {Beniamini}, Andrey {Timokhin}, Matthew~G. {Baring}, Alexander~J. {van der Horst}, Alice~K. {Harding}, and Demosthenes {Kazanas}.
\newblock {The Fast Radio Burst Luminosity Function and Death Line in the Low-twist Magnetar Model}.
\newblock {\em \apj}, 891(1):82, March 2020.

\bibitem{2020ApJ...903L..38W}
Zorawar {Wadiasingh} and Cecilia {Chirenti}.
\newblock {Fast Radio Burst Trains from Magnetar Oscillations}.
\newblock {\em \apjl}, 903(2):L38, November 2020.

\bibitem{2023HEAD...2011649W}
Zorawar {Wadiasingh}, Constantinos {Kalapotharakos}, Matthew {Baring}, Kun {Hu}, George {Younes}, Alice {Harding}, Ersin {Gogus}, and Chryssa {Kouveliotou}.
\newblock {Strong Lensing in Magnetar Burst Fireballs: Transfer Functions}.
\newblock In {\em AAS/High Energy Astrophysics Division}, volume~55 of {\em AAS/High Energy Astrophysics Division}, page 116.49, September 2023.

\bibitem{2019ApJ...879....4W}
Zorawar {Wadiasingh} and Andrey {Timokhin}.
\newblock {Repeating Fast Radio Bursts from Magnetars with Low Magnetospheric Twist}.
\newblock {\em \apj}, 879(1):4, July 2019.

\bibitem{2008nops.book...19W}
Christopher~W. {Walter}.
\newblock {The Super-Kamiokande Experiment}.
\newblock In {\em Neutrino Oscillations: Present Status and Future Plans. Edited by THOMAS JENNIFER A \& VAHLE PATRICIA L. Published by World Scientific Publishing Co. Pte. Ltd}, pages 19--43. 2008.

\bibitem{2006ApJ...637L.117W}
Anna~L. {Watts} and Tod~E. {Strohmayer}.
\newblock {Detection with RHESSI of High-Frequency X-Ray Oscillations in the Tailof the 2004 Hyperflare from SGR 1806-20}.
\newblock {\em \apjl}, 637(2):L117--L120, February 2006.

\bibitem{wei21ApJ}
Jun-Jie {Wei}, Xue-Feng {Wu}, Zi-Gao {Dai}, Fa-Yin {Wang}, Pei {Wang}, Di~{Li}, and Bing {Zhang}.
\newblock {Similar Scale-invariant Behaviors between Soft Gamma-Ray Repeaters and an Extreme Epoch from FRB 121102}.
\newblock {\em \apj}, 920(2):153, October 2021.

\bibitem{Woods2007}
Peter~M. {Woods}, Chryssa {Kouveliotou}, Mark~H. {Finger}, Ersin {G{\"o}{\v{g}}{\"u}{\c{s}}}, Colleen~A. {Wilson}, Sandeep~K. {Patel}, Kevin {Hurley}, and Jean~H. {Swank}.
\newblock {The Prelude to and Aftermath of the Giant Flare of 2004 December 27: Persistent and Pulsed X-Ray Properties of SGR 1806-20 from 1993 to 2005}.
\newblock {\em \apj}, 654(1):470--486, January 2007.

\bibitem{2023arXiv231214833Y}
Yi-Han~Iris {Yin}, Zhao~Joseph {Zhang}, Jun {Yang}, Run-Chao {Chen}, Umer {Rehman}, {Varun}, and Bin-Bin {Zhang}.
\newblock {A Comptonized Fireball Bubble Fits the Second Extragalactic Magnetar Giant Flare GRB 231115A}.
\newblock {\em arXiv e-prints}, page arXiv:2312.14833, December 2023.

\bibitem{younes23NatAs}
G.~{Younes}, M.~G. {Baring}, A.~K. {Harding}, T.~{Enoto}, Z.~{Wadiasingh}, A.~B. {Pearlman}, W.~C.~G. {Ho}, S.~{Guillot}, Z.~{Arzoumanian}, A.~{Borghese}, K.~{Gendreau}, E.~{G{\"o}{\v{g}}{\"u}{\c{s}}}, T.~{G{\"u}ver}, A.~J. {van der Horst}, C.~P. {Hu}, G.~K. {Jaisawal}, C.~{Kouveliotou}, L.~{Lin}, and W.~A. {Majid}.
\newblock {Magnetar spin-down glitch clearing the way for FRB-like bursts and a pulsed radio episode}.
\newblock {\em Nature Astronomy}, 7:339--350, March 2023.

\bibitem{younes2021NatAs}
G.~{Younes}, M.~G. {Baring}, C.~{Kouveliotou}, Z.~{Arzoumanian}, T.~{Enoto}, J.~{Doty}, K.~C. {Gendreau}, E.~{G{\"o}{\v{g}}{\"u}{\c{s}}}, S.~{Guillot}, T.~{G{\"u}ver}, A.~K. {Harding}, W.~C.~G. {Ho}, A.~J. {van der Horst}, C.~P. {Hu}, G.~K. {Jaisawal}, Y.~{Kaneko}, B.~J. {LaMarr}, L.~{Lin}, W.~{Majid}, T.~{Okajima}, J.~{Pope}, P.~S. {Ray}, O.~J. {Roberts}, M.~{Saylor}, J.~F. {Steiner}, and Z.~{Wadiasingh}.
\newblock {Broadband X-ray burst spectroscopy of the fast-radio-burst-emitting Galactic magnetar}.
\newblock {\em Nature Astronomy}, 5:408--413, April 2021.

\bibitem{younes2016ApJ}
G.~{Younes}, C.~{Kouveliotou}, O.~{Kargaltsev}, R.~{Gill}, J.~{Granot}, A.~L. {Watts}, J.~{Gelfand}, M.~G. {Baring}, A.~{Harding}, G.~G. {Pavlov}, A.~J. {van der Horst}, D.~{Huppenkothen}, E.~{G{\"o}{\u{g}}{\"u}{\c{s}}}, L.~{Lin}, and O.~J. {Roberts}.
\newblock {The Wind Nebula around Magnetar Swift J1834.9-0846}.
\newblock {\em \apj}, 824(2):138, June 2016.

\bibitem{2014ApJ...785...52Y}
G.~{Younes}, C.~{Kouveliotou}, A.~J. {van der Horst}, M.~G. {Baring}, J.~{Granot}, A.~L. {Watts}, P.~N. {Bhat}, A.~{Collazzi}, N.~{Gehrels}, N.~{Gorgone}, E.~{G{\"o}{\u{g}}{\"u}{\c{s}}}, D.~{Gruber}, S.~{Grunblatt}, D.~{Huppenkothen}, Y.~{Kaneko}, A.~{von Kienlin}, M.~{van der Klis}, L.~{Lin}, J.~{Mcenery}, T.~{van Putten}, and R.~A.~M.~J. {Wijers}.
\newblock {Time Resolved Spectroscopy of SGR J1550-5418 Bursts Detected with Fermi/Gamma-Ray Burst Monitor}.
\newblock {\em \apj}, 785(1):52, April 2014.

\bibitem{younes17ApJ:1806}
George {Younes}, Matthew~G. {Baring}, Chryssa {Kouveliotou}, Alice {Harding}, Sophia {Donovan}, Ersin {G{\"o}{\u{g}}{\"u}{\c{s}}}, Victoria {Kaspi}, and Jonathan {Granot}.
\newblock {The Sleeping Monster: NuSTAR Observations of SGR 1806-20, 11 Years After the Giant Flare}.
\newblock {\em \apj}, 851(1):17, December 2017.

\bibitem{2020ApJ...904L..21Y}
George {Younes}, Tolga {G{\"u}ver}, Chryssa {Kouveliotou}, Matthew~G. {Baring}, Chin-Ping {Hu}, Zorawar {Wadiasingh}, Beste {Begi{\c{c}}arslan}, Teruaki {Enoto}, Ersin {G{\"o}{\u{g}}{\"u}{\c{s}}}, Lin {Lin}, Alice~K. {Harding}, Alexander~J. {van der Horst}, Walid~A. {Majid}, Sebastien {Guillot}, and Christian {Malacaria}.
\newblock {NICER View of the 2020 Burst Storm and Persistent Emission of SGR 1935+2154}.
\newblock {\em \apjl}, 904(2):L21, December 2020.

\bibitem{younes22ApJ:1830}
George {Younes}, Samuel~K. {Lander}, Matthew~G. {Baring}, Teruaki {Enoto}, Chryssa {Kouveliotou}, Zorawar {Wadiasingh}, Wynn C.~G. {Ho}, Alice~K. {Harding}, Zaven {Arzoumanian}, Keith {Gendreau}, Tolga {G{\"u}ver}, Chin-Ping {Hu}, Christian {Malacaria}, Paul~S. {Ray}, and Tod~E. {Strohmayer}.
\newblock {Pulse Peak Migration during the Outburst Decay of the Magnetar SGR 1830-0645: Crustal Motion and Magnetospheric Untwisting}.
\newblock {\em \apjl}, 924(2):L27, January 2022.

\bibitem{younes20ApJ:2259}
George {Younes}, Paul~S. {Ray}, Matthew~G. {Baring}, Chryssa {Kouveliotou}, Corinne {Fletcher}, Zorawar {Wadiasingh}, Alice~K. {Harding}, and Adam {Goldstein}.
\newblock {A Radiatively Quiet Glitch and Anti-glitch in the Magnetar 1E 2259+586}.
\newblock {\em \apjl}, 896(2):L42, June 2020.

\bibitem{2022hxga.book...86Y}
Weimin {Yuan}, Chen {Zhang}, Yong {Chen}, and Zhixing {Ling}.
\newblock {The Einstein Probe Mission}.
\newblock In {\em Handbook of X-ray and Gamma-ray Astrophysics}, page~86. 2022.

\bibitem{2014PhRvD..89j7303Z}
Bei {Zhou}, Xiang {Li}, Tao {Wang}, Yi-Zhong {Fan}, and Da-Ming {Wei}.
\newblock {Fast radio bursts as a cosmic probe?}
\newblock {\em \prd}, 89(10):107303, May 2014.

\bibitem{2023SciA....9F6198Z}
Weiwei {Zhu}, Heng {Xu}, Dejiang {Zhou}, Lin {Lin}, Bojun {Wang}, Pei {Wang}, Chunfeng {Zhang}, Jiarui {Niu}, Yutong {Chen}, Chengkui {Li}, Lingqi {Meng}, Kejia {Lee}, Bing {Zhang}, Yi~{Feng}, Mingyu {Ge}, Ersin {G{\"o}{\u{g}}{\"u}{\c{s}}}, Xing {Guan}, Jinlin {Han}, Jinchen {Jiang}, Peng {Jiang}, Chryssa {Kouveliotou}, Di~{Li}, Chenchen {Miao}, Xueli {Miao}, Yunpeng {Men}, Chenghui {Niu}, Weiyang {Wang}, Zhengli {Wang}, Jiangwei {Xu}, Renxin {Xu}, Mengyao {Xue}, Yuanpei {Yang}, Wenfei {Yu}, Mao {Yuan}, Youling {Yue}, Shuangnan {Zhang}, and Yongkun {Zhang}.
\newblock {A radio pulsar phase from SGR J1935+2154 provides clues to the magnetar FRB mechanism}.
\newblock {\em Science Advances}, 9(30):eadf6198, July 2023.

\end{thebibliography}

\end{document}